\documentclass[twocolumn]{aastex62}

\usepackage{color}
\usepackage{dsfont}
\usepackage{physics}
\usepackage{amsmath, amssymb}

\newcommand{\cmnt}[1]{}

\newcommand{\tj}[6]{ \begin{array}{ccc}
   #1 & ~#2~ & #3 \\
   #4 & ~#5~ & #6 
\end{array}}

\newcommand{\sj}[9]{ \begin{array}{ccc}
   #1 & ~#2~ & #3 \\
   #4 & ~#5~ & #6 \\
   #7 & ~#8~ & #9
\end{array}}
\submitjournal{ApJ}

\shorttitle{}
\shortauthors{}

\begin{document}

\title{Detecting Magnetic Fields in Exoplanets with Spectropolarimetry of the Helium Line at 1083 nm}

\correspondingauthor{Antonija Oklop\v ci\'c}
\email{antonija.oklopcic@cfa.harvard.edu}

\author[0000-0002-9584-6476]{Antonija Oklop\v ci\'c}\altaffiliation{NHFP Sagan Fellow}
\affil{Center for Astrophysics $\vert$ Harvard \& Smithsonian, 60 Garden Street, MS-16, Cambridge, MA 02138, USA}
\author[0000-0002-1717-4424]{Makana Silva}
\affil{Center for Cosmology and AstroParticle Physics (CCAPP), The Ohio State University, Columbus, Ohio 43210, USA}
\affil{Department of Physics, The Ohio State University, 191 West Woodruff Avenue, Columbus, Ohio 43210, USA}
\author[0000-0002-6998-6678]{Paulo Montero-Camacho}
\affil{Center for Cosmology and AstroParticle Physics (CCAPP), The Ohio State University, Columbus, Ohio 43210, USA}
\affil{Department of Physics, The Ohio State University, 191 West Woodruff Avenue, Columbus, Ohio 43210, USA}
\affil{Department of Astronomy, Tsinghua Center for Astrophysics, Tsinghua University, Beijing 100084, China}
\author[0000-0002-2951-4932]{Christopher M. Hirata}
\affil{Center for Cosmology and AstroParticle Physics (CCAPP), The Ohio State University, Columbus, Ohio 43210, USA}
\affil{Department of Physics, The Ohio State University, 191 West Woodruff Avenue, Columbus, Ohio 43210, USA}
\affil{Department of Astronomy, The Ohio State University, 140 West 18th Avenue, Columbus, Ohio 43210, USA}

\begin{abstract}
The magnetic fields of the solar system planets provide valuable insights into the planets' interiors and can have dramatic consequences for the evolution of their atmospheres and interaction with the solar wind. However, we have little direct knowledge of magnetic fields in exoplanets. Here we present a method for detecting magnetic fields in the atmospheres of close-in exoplanets based on spectropolarimetric transit observations at the wavelength of the helium line at 1083 nm. This methodology has been successfully applied before for exploring magnetic fields in solar coronal filaments. Strong absorption signatures (transit depths on the order of a few percent) in the 1083 nm line have recently been observed for several close-in exoplanets. We show that in the conditions in these escaping atmospheres, metastable helium atoms should be optically pumped by the starlight and, for field strengths more than a few$\times 10^{-4}$ G, should align with the magnetic field. This results in linearly polarized absorption at 1083 nm that traces the field direction (the Hanle effect), which we explore by both analytic computation and with the {\sc Hazel} numerical code. The linear polarization $\sqrt{Q^2+U^2}/I$ ranges from $\sim 10^{-3}$ in optimistic cases down to a few$\times 10^{-5}$ for particularly unfavorable cases, with very weak dependence on field strength. The line-of-sight component of the field results in a slight circular polarization (the Zeeman effect), also reaching $V/I\sim {\rm few}\times 10^{-5}(B_\parallel/10\,{\rm G})$. We discuss the detectability of these signals with current (SPIRou) and future (extremely large telescope) high-resolution infrared spectropolarimeters, and we briefly comment on possible sources of astrophysical contamination.
\end{abstract}


\section{Introduction} 
\label{sec:intro}

One of the most important planetary properties is the presence or absence of a global magnetic field. Since the magnetic field dynamos rely on differential rotation or convection of an electrically conducting fluid in the planet's interior, detecting and measuring planetary magnetic fields is one of the few ways to obtain information about the structure, composition, and dynamics of planetary interiors. Furthermore, the presence of a global magnetic field---shielding the planet from impacts of high-energy particles in the stellar wind---can have important consequences for the extent and longevity of a planetary atmosphere and ultimately determine whether a planet is habitable or not.

 It is still an open question whether magnetic field dynamos in planets, brown dwarfs, and stars all share the same physical origin. Magnetic field strengths on the order of a kilogauss have been detected in brown dwarfs and low-mass stars \citep[e.g.][]{ReinersBasri2007, Morin2010, Kao2018}. Most planets and some satellites in the solar system have or had in their past a global magnetic field, with average magnetic field strengths at their surface up to $\sim 5$~G \citep[in the case of Jupiter;][]{SchubertSoderlund2011}. One of the main challenges for understanding planetary dynamos is the small sample of solar system planets. Measuring the properties of magnetic fields in exoplanets could expand that sample and provide valuable insights into the origin of planetary magnetic fields and its importance for planetary evolution. 
 
 One of the most promising avenues for detecting magnetic fields in exoplanets is using the radio electron cyclotron maser emission \citep[e.g.][]{Hallinan2008}. However, no direct detections of magnetic fields in exoplanets have been made so far \citep[e.g.][]{Murphy2015, Lazio2018}. Indirect detection of magnetic fields in several hot Jupiters, obtained by modeling star--planet magnetic interaction and its effect on stellar chromospheric emission, has recently been reported by \citet{Cauley2019}.

Magnetic fields in stars, including the Sun, can be directly observed by their effect on atoms and molecules in stellar atmospheres. By changing the structure of atomic and molecular energy levels, magnetic fields affect the spectral line profiles and polarization properties of stellar radiation \citep[e.g.][]{Landi2004}. Most easily observable manifestations of magnetic fields arise from the Zeeman effect, which causes magnetic sublevels of a given atomic state to split in energy. As a result, the spectral line appears broadened or split into multiple components, which are polarized. By observing radiation polarization in sunspots, \citet{Hale1908} was the first to infer the presence of a magnetic field on the Sun, and most of our current knowledge of stellar magnetism comes from Zeeman spectroscopy and spectropolarimetry \citep[e.g.][]{DonatiLandstreet2009}. 

Polarization in spectral lines, however, can also be induced by anisotropic radiation pumping, which creates population imbalances between different atomic sublevels and hence linear polarization in the emergent radiation \citep[e.g., see the monograph by][]{Landi2004}. In the presence of a magnetic field, this atomic-level polarization gets modified through the action of the Hanle effect \citep{Hanle1924}. This effect is sensitive to much weaker magnetic fields compared to the Zeeman effect, and its diagnostic potential has been utilized by the solar physics community for many years \citep[e.g.][]{Leroy1977,Stenflo1998, TrujilloBueno2002, TrujilloBueno2005}. Of particular interest is the paper by \citet{TrujilloBueno2002}, which shows that in prominences and filaments levitating in the solar corona, the metastable lower level of the He {\sc i} 1083 nm triplet is significantly polarized by the anisotropic radiation from the underlying solar disk, and that a significant amount of linear polarization is produced by selective absorption processes \citep[see also][]{TrujilloBueno2007}. Interestingly, as pointed out by \citet{TrujilloBueno2002}, the mere detection of He {\sc i} 1083 nm linear polarization when observing filaments at the solar disk center implies the presence of a magnetic field, inclined with respect to the line of sight, in the filament plasma. 

Founded on the same physical principles, here we propose a similar method for detecting magnetic fields in exoplanets. The method is based on high-resolution transmission spectropolarimetry in the helium absorption line at 1083 nm; it is applicable to transiting hot planets with extended or escaping atmospheres. Observing a linear polarization signal in this spectral line during an exoplanet transit could reveal the presence of a magnetic field in the planet's atmosphere. The method is sensitive to a broad range of magnetic field strengths, including the field strengths found in the solar system planets. This paper is structured as follows. In \S\ref{sec:background} we provide the basic background on the helium line at 1083~nm and the polarimetry of spectral lines in the presence of magnetic fields. In \S\ref{sec:analytic} we present an analytic calculation of the effect an external magnetic field has on the polarization of the helium line. In \S\ref{sec:numerical} we present the results of numerical calculations for different magnetic field strengths and geometries using the \textsc{hazel} code \citep{AsensioRamos2008}. In \S\ref{sec:discussion} we examine the prospects for observing the calculated polarization signals and discuss possible sources of contamination, and in \S\ref{sec:summary} we summarize our results.

\section{Background}\label{sec:background}

\subsection{Helium Absorption Line at 1083 nm}

The helium line triplet at 1083~nm has recently been established as a powerful new diagnostic of upper atmospheres of exoplanets. An absorption signature in this line produced by a transiting exoplanetary atmosphere is sensitive to the physical properties (i.e. gas temperature, density, composition) and the dynamics (i.e. winds and outflows) of the atmospheric layers extending to distances of a few planetary radii, that is, the thermosphere and exosphere \citep{OklopcicHirata2018}. Studying this region of the atmosphere, close to the planet's Roche radius, is important for understanding the process of atmospheric escape and its effect on planetary evolution. 

Helium absorption at 1083~nm was first detected in transit spectroscopy of WASP-107b by \citet{Spake2018}, using the low-resolution data from the \textit{Hubble Space Telescope}/Wide Field Camera 3. Since then, spectrally resolved absorption signatures of helium have been obtained for several exoplanets (HAT-P-11b, WASP-69b, HD~189733b, WASP-107b, HD~209458b) using high-resolution ground-based observations \citep{Allart2018, Nortmann2018, Salz2018, Allart2019, Alonso-Floriano2019}.

\begin{figure}
\centering
\includegraphics[width=0.49\textwidth]{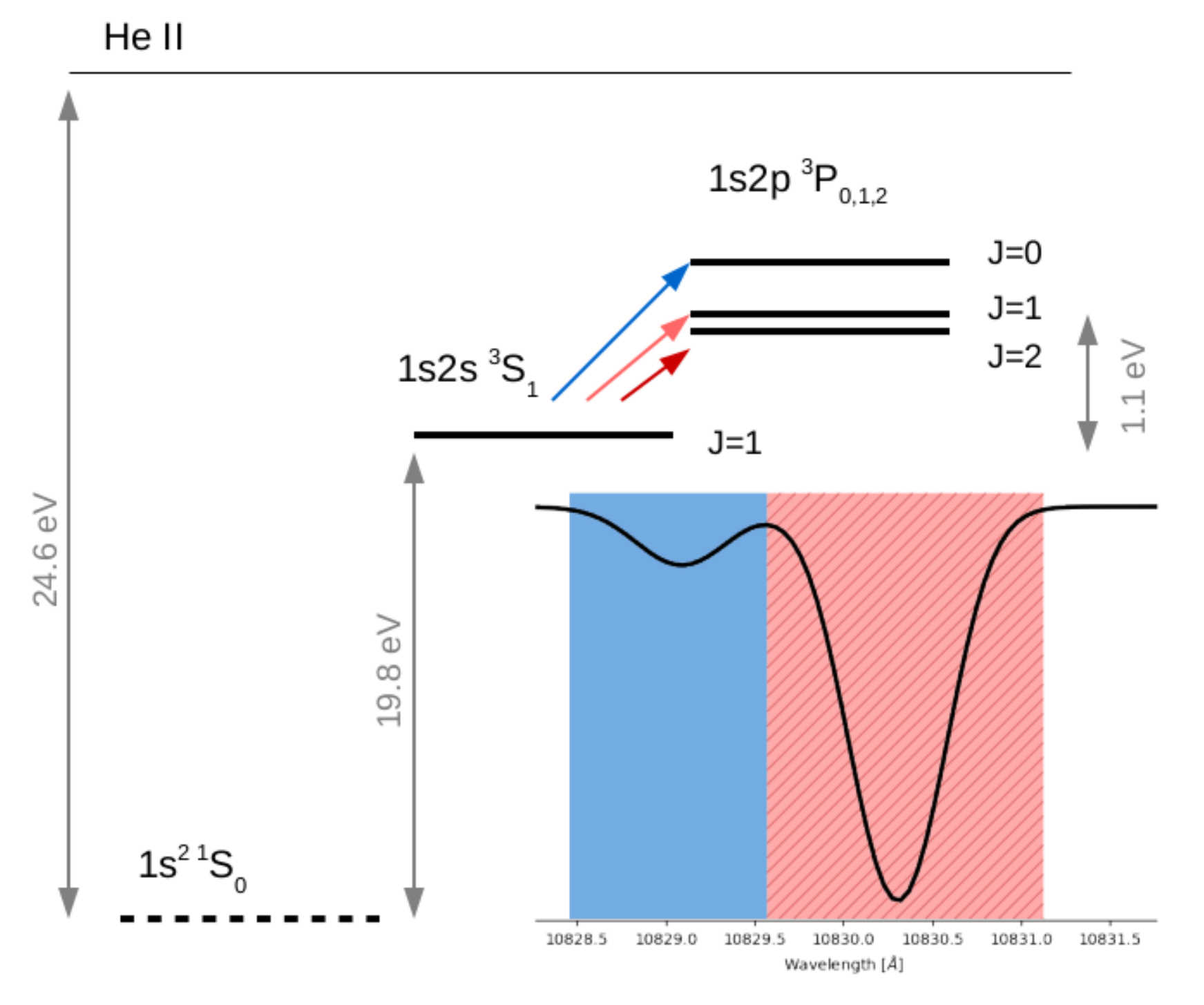}
\caption{Schematic representation of atomic levels involved in the helium 1083~nm transitions (solid lines). The transition between the metastable (2$^3$S) level and the $J=0$ level of 2$^3$P produces the ``blue component'' of the 1083~nm absorption line triplet, whereas the other two transitions blend together to form the ``red component.'' Energy separations are not drawn to scale.}
\label{fig:atomic_levels}
\end{figure}

The absorption triplet at 1083~nm originates from an excited 2$^3$S$_1$ level of the neutral helium atom in the triplet configuration. This level has only highly forbidden decays to the helium ground state (1$^1$S$_0$), and hence it is metastable, allowing a significant population of helium atoms in this excited level to build up in planetary thermospheres and exospheres. The upper level of the 1083~nm transition (2$^3$P$_{2,1,0}$) is split by fine structure into three levels with quantum numbers $J=2$, $J=1$, and $J=0$. These three levels correspond to lines with wavelengths (in air) of 1083.034~nm, 1083.025~nm, and 1082.909~nm, respectively \citep{NIST_ASD}. The first two are practically indistinguishable, and we refer to them jointly as the ``red component,'' whereas the third, the ``blue component,'' is usually separated. A schematic representation of the atomic transitions relevant for the helium 1083~nm line triplet and the line components they produce is given in \autoref{fig:atomic_levels}. 

\subsection{Polarization of Absorption Lines in the Presence of a Magnetic Field} 
\label{ss:specpol}

\begin{figure}
    \centering
    \includegraphics{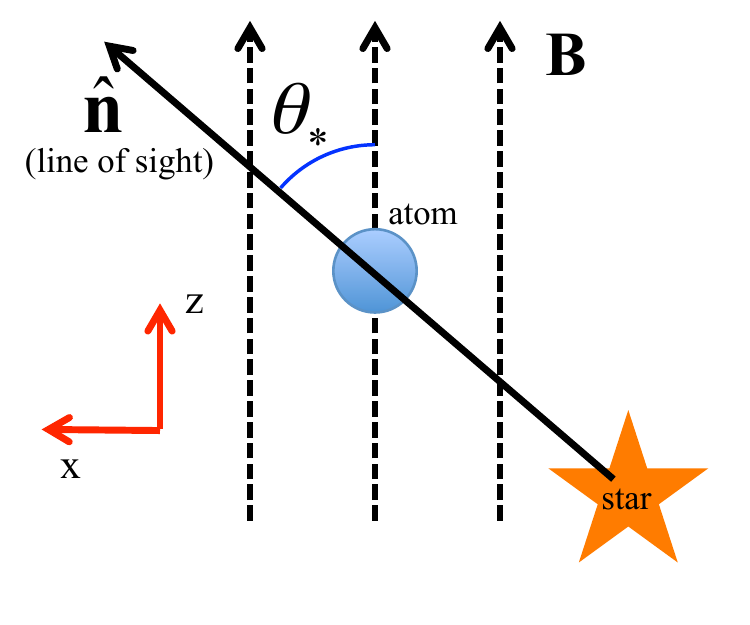}
    \caption{Setup for \S\ref{ss:specpol}.}
    \label{fig:overview}
\end{figure}

\begin{figure}
\centering
\includegraphics[width=0.47\textwidth]{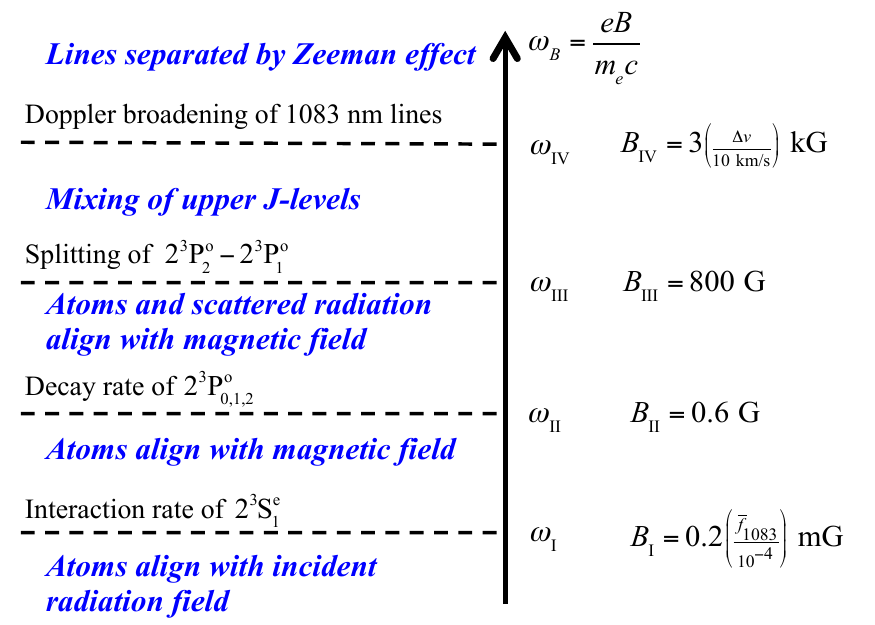}
\caption{Different regimes of magnetic field strength relevant to the \ion{He}1 1083 nm triplet. At very low field strength, $B<B_{\rm I}$, the precession rate is less than the interaction rate of \ion{He}1 2$^3$S, and the magnetic field is only a small perturbation on the alignment of the atoms. When $B>B_{\rm I}$, the precession rate is fast, and the metastable helium atoms have an axisymmetric spin distribution around the magnetic field. At even higher fields, $B>B_{\rm II}$, the precession rate is faster than the decay rate of 2$^3$P, and the scattered 1083 nm photons (emitted when the atoms decay back to 2$^3$S) are also axisymmetric around ${\boldsymbol B}$. At $B>B_{\rm III}$, the precession rate exceeds the spin-orbit splitting 2$^3$P$^{\rm o}_2$--2$^3$P$^{\rm o}_1$, and the upper $J$-levels are mixed. Finally, at very large fields, $B>B_{\rm IV}$, the Zeeman splitting exceeds the Doppler width of the line, and the components of different polarization are separated. Note $B_{\rm II}$ and $B_{\rm III}$ are fixed by atomic physics, but $B_{\rm I}$ and $B_{\rm IV}$ depend on the environment; we show representative values. The treatment in \S\ref{sec:analytic} assumes $B_{\rm I}<B<B_{\rm III}$.}
\label{fig:Bfield}
\end{figure}

We now review how atoms placed in a magnetic field and subject to an incident radiation field can become aligned\footnote{Atoms are \textit{aligned} with the magnetic field if the distribution of their spins is anisotropic but approximately axisymmetric around the field. Quantum mechanically, this means the density matrix is not proportional to the identity, but is approximately diagonal in the $|M_J\rangle$ basis with the $z$-axis along the field.} with the field and exhibit polarization-selective absorption. The setup is shown in \autoref{fig:overview}. This effect is well studied in the context of solar physics \citep[see][]{Trujillo1997, TrujilloBueno2002}, and here we will argue that all the circumstances required for a spectral line to be a useful probe of magnetic fields are satisfied by the \ion{He} 1 1083 nm triplet in extended exoplanet atmospheres around late-type stars (with the possible exception of the magnetic field strength, which is not known at present, and which we hope to constrain). Along the way, we will discuss the relevant field strengths, both in terms of $B$ and in terms of the cyclotron frequency $\omega_B = eB/m_ec$, which to an order of magnitude is the precession frequency of an atom in a magnetic field.\footnote{The precession frequency of an atom is $\omega_{\rm prec} = \frac{g_J}2\omega_B$, where $g_J$ is the Land\'e $g$-factor. Standard formulae (e.g., \citealt{RybickiLightman}, Eq.~9.35b) give $\frac{g_J}2=1$ for 2$^3$S and $\frac{g_J}2=\frac34$ for 2$^3$P.} Key field strengths where there is a qualitative change in behavior, $B_{\rm I}$, $B_{\rm II}$, $B_{\rm III}$, and $B_{\rm IV}$, are shown in \autoref{fig:Bfield}.

The first requirement is that the lower level of the transition have a nonzero angular momentum $J_l$ so that there is something to align. If we want linearly polarized absorption, the atom must be able to have a spin 2 component of the density matrix (since $Q$ and $U$ have spin 2), so this requires $J_l\ge 1$ by the triangle inequality. This is the case for the level 2$^3$S$^{\rm e}_1$, which has $J_l=1$.

Second, absorption of radiation followed by emission must be able to align the atoms; that is, in the scattering process
\begin{equation}
{\rm He}(2^3{\rm S}^e_1) + \gamma \rightarrow {\rm He}(2^3{\rm P}^o_{0,1,2}) \rightarrow {\rm He}(2^3{\rm S}^e_1) + \gamma,
\label{eq:scatproc}
\end{equation}
we must be able to start from an initial random state ($M_J=-1$, 0, or $+1$ equally likely) and leave the atom in a nonrandom final state. Here in the absorption step, the angular momentum of the photon is transferred to the orbital angular momentum of the electrons, and in the emission step the orbital angular momentum of the electrons is transferred to the outgoing photon. If there were no spin-orbit coupling, then the total electron spin would be conserved through this whole process, and the final $M_J$ would equal the initial $M_J$. However, the fine structure splittings of the \ion{He}1 2$^3$P$^{\rm o}_{0,1,2}$ levels are large compared to their intrinsic width, which means that the total electron spin ${\bf S}$ is {\em not} conserved; rather, in the intermediate state, the spin and orbital angular momenta can precess around each other for many cycles before the atom decays back to 2$^3$S$^{\rm e}_1$. Therefore, the final spin of the atom will have some dependence on the angular momentum of the incident photon. Since most incident photons come from the star and carry angular momentum $\pm\hbar$ (but not zero) projected along the line-of-sight ($\hat{\bf n}$) direction, this final spin is not random. The distribution of final states will depend on the angle $\theta_\star$ between the line of sight and the magnetic field. If the field strength is small enough to ignore Zeeman splittings (see below), then the alignment of atoms will be of the quadrupole or ``headless vector'' nature; that is, the $M_J=-1$ and $M_J=+1$ final states are equally likely, but their probability will differ from $M_J=0$.

Third, the atoms must precess around the magnetic field fast enough that they align with the field and not with the direction of incident radiation (in the latter case---the \textit{zero-field case}, according to the terminology used in \citealt{Landi2004}---we know from symmetry considerations that they will produce no polarization). This means that the precession rate $\sim \omega_B$ must be large compared to the rate at which the atoms scatter photons, $3A_{ul}\bar f_{1083}$, where 3 is a ratio of statistical weights and $\bar f_{1083}$ is the isotropically averaged phase space density of photons in the 1083 nm line.\footnote{This is related to the isotropically averaged specific intensity by $J_\nu = (2h\nu^3/c^2)\bar f_{1083}$.} This requirement implies $B>B_{\rm I} = 0.2(\bar f_{1083}/10^{-4})\,$mG, where $\bar f_{1083}$ is typically of order $10^{-4}$ near hot exoplanets. This is the lower-level Hanle effect regime.

Fourth, the atoms in the lower level must not be depolarized or depopulated by other interactions. This means that the scattering rate $3A_{ul}\bar f_{1083}$ in the 1083 nm triplet must be large compared to the rate of spontaneous decay $A_l$, collisions $\sum_i n_i\langle \sigma v\rangle_i$ (including all available projectiles $i$), and photoionization $\Phi_l$. Here we are helped by the metastable nature of the 2$^3$S$^{\rm e}_1$ level (small $A_l$), the high radiation flux and low gas density in the escaping atmospheres (see the \autoref{sec:appendix}), and the red spectrum of the K-type stars hosting \ion{He}1 detections, which results in a much higher rate for 1083 nm scattering than photoionization from the 2$^3$S$^{\rm e}_1$ level (requiring $\lambda<260$ nm photons). A review of most of these rates can be found in Section 3.3 of \citet{OklopcicHirata2018} and in Figure 4 of \citet{Oklopcic19}.

Finally, it is not enough to align the spins of the atoms; we must also have a spectral diagnostic that is sensitive to this alignment. The key to this is the fine structure splitting: the cross section for $2^3{\rm S}^{\rm e}_1-2^3{\rm P}^{\rm o}_J$ for a given final $J$ depends on the relative orientations of the initial spin of the atom and the polarization of the incoming light. This is simplest to understand for the absorption line leading to the $J=0$ upper level, which always has $M_J=0$. The transition obeys a selection rule in which the $E_z$ component of the incident electric field (which preserves symmetry around the $z$-axis) must lead to $\Delta M_J=0$ by conservation of angular momentum, whereas the $E_x$ and $E_y$ components lead to $\Delta M_J=\pm 1$. Thus the $M_J=0$ initial state only absorbs light in the $2^3{\rm S}^{\rm e}_1-2^3{\rm P}^{\rm o}_0$ line for the vertical polarization (see \autoref{fig:overview}), whereas the $M_J=\pm 1$ initial states can absorb light in the horizontal polarization as well (at $\theta_\star=\pi/2$, this must be horizontal, but in general it may be either vertical or horizontal). Similar but more complicated considerations apply to the $J=1$ and $J=2$ upper levels. This means that the ``headless vector'' alignment of helium atoms -- where $M_J=0$ has a different occupation from $M_J=\pm 1$ -- should produce linearly polarized absorption.

At high field strengths, other effects begin to occur. For $B>B_{\rm II}$ (see \autoref{fig:Bfield}) or $\omega_B>A_{ul}$, the excited \ion{He}1 2$^3$P atoms can precess before they decay (i.e. the upper-level Hanle effect); in this case, the scattered radiation will be randomized in longitude around the magnetic field. However, because in transits we observe absorption and not primarily the scattered light, this should have only a minor influence on the observed transit depth. At $B>B_{\rm III}$ or $\omega_{\rm B}>\Delta E(2^3{\rm P}^{\rm o}_2-2^3{\rm P}^{\rm o}_1)/\hbar$, there is mixing of the upper $J$-levels; this changes the calculation of the optical pumping, but the alignment mechanism still works. At very high magnetic field strengths (i.e. in the \textit{intense field regime}) there will be a Zeeman splitting of the 1083 nm line components that can result in circular polarization. The lines become separated for $\omega_B>\omega_0(\Delta v/c)$ or $B>B_{\rm IV} = 3(\Delta v/10\,{\rm km}\,{\rm s}^{-1})\,$kG, where $\omega_0 = 1.74\times 10^{15}\,{\rm s}^{-1}$ is the angular frequency of the line. Even for smaller field strengths---in the so-called \textit{intermediate regime}---one expects the line components with different circular polarization to be slightly separated, such that Stokes $V$ has a negative--positive pattern around each line. We do not consider this in our simplified analytic calculation in \S\ref{sec:analytic}, but it is included in the numerical calculations of \S\ref{sec:numerical} using {\sc Hazel}.

\section{Simplified Analytic Calculation}
\label{sec:analytic}

In this section, we will go through a simplified calculation for obtaining the optical depth and Stokes parameters \textit{Q} and \textit{I} for transmitted radiation from metastable helium atoms in a magnetic field that are pumped by incident starlight. In \S \ref{sec:assumptions}, we state our underlying assumptions. We set up the optical pumping calculation in \S\ref{sec:calc} and solve the resulting equations for the atomic density matrix in \S\ref{sec:calc-sol}. We arrive at the analytic results for polarization-selective absorption in \S\ref{sec:analytic_results}.

\subsection{Assumptions, Range of Validity, and Notation}
\label{sec:assumptions}

In order to calculate an analytic solution to this problem, we first establish our assumptions about the system of interest. We first simplify the problem by assuming a uniform magnetic field and a small optical depth (so that the probability of multiple scattering in the 1083 nm line is small). These approximations are purely to simplify the problem and estimate the order of magnitude of the signal strength. For example, inhomogeneous magnetic fields could be modeled by superposing several field geometries; scattered light adds complexity and can be treated with numerical codes as in \S\ref{sec:numerical}, but does not change the essential aspects of the problem.

In treating the pumping of the metastable helium atoms by the ambient 1083 nm radiation field, we work in the case where the angular size of the star is negligible (so that $\cos\theta$ does not vary over the star's disk, and we can set $\theta\approx\theta_\star$ for incoming radiation) and the emitted light is of a smooth spectrum and unpolarized. This is justified in that the planet is many stellar radii from the star, and the stellar \ion{He}1 1083 nm lines are weaker than the continuum. Similarly, if we are at least a couple planetary radii from the planet, we neglect 1083 nm photons from the star that reflect off the planet.

The next assumption is the range of the magnetic field strength. As stated in \S\ref{ss:specpol}, we want a magnetic field with strength great enough to allow for fast precession of the atom but not strong enough to mix the $J$-levels in 2$^3$P$^{\rm o}_{2,1,0}$. This puts $B_{\rm min} \approx B_{\rm I} = 0.2 \ \textup{mG}$ and $B_{\rm max} \approx B_{\rm III} = 800 \ \textup{G}$. Fast precession guarantees that the density matrix of metastable helium \ion{He}1 2$^3$S$^{\rm e}_1$ is diagonal in the $M_J$ basis.

When describing states, we will use capital letters $N, R, T$ and $W$ to denote states in the 2$^3$S$^{\rm e}_1$ metastable level and Greek letters $\mu$ and $\nu$ to denote states in the 2$^3$P$^{\rm o}_{2,1,0}$ excited levels. Components of vectors will be written with lowercase letters. Quantum numbers of these states will be written with the state as a subscript, such as in $J_\mu$ and $M_{J,\mu}$, where $J$ and $M_J$ follow standard atomic physics notation.

\subsection{Polarization of the Metastable Helium Atoms: Setup}
\label{sec:calc}

In order to compare the analytic solution to the results of the {\sc Hazel} code, we will want to compute the polarization of the transmitted light under the presence of a uniform magnetic field. Our first step is to calculate the density matrix of the metastable \ion{He}1 atoms. This is a $3\times 3$ Hermitian matrix, with trace equal to 1. We want to find steady-state solutions to the density matrix evolution equation for helium atoms in the $2^3{\rm S}^{\rm e}_1$ level:
\begin{equation}
\label{eq:steady}
 \dot{\rho}_{B,TW} + \dot{\rho}_{{\rm depop},TW} + \dot{\rho}_{{\rm repop},TW} = 0 \,.
\end{equation}
The first term is given by the Zeeman splitting of states $T$ and $W$:
\begin{eqnarray}
\label{eq:B}
 \dot{\rho}_{B,TW} = i \frac{g_l}2 \omega_B (M_{J,W}-M_{J,T})\rho_{TW},
 \end{eqnarray} 
where the Land\'e $g$-factor for the lower level is $g_l=2$. The second term has the following form
 \begin{eqnarray}
 \label{eq:depop}
 \dot{\rho}_{{\rm depop},TW} = -\frac{3}{4\pi}A_{ul}f_{\rm star}(\omega_{0})\Omega_{\rm star}\rho_{TW},
\end{eqnarray}
where $A_{ul}$ is the Einstein coefficient for spontaneous emission, and $f_{\rm star}(\omega)$ and $\Omega_{\rm star}$ are the phase space of incoming photons from the star and the solid angle of the star as seen by the atom, respectively. (The angular average phase space density is $\bar f_{1083} = f_{\rm star}(\omega_0) \times \Omega_{\rm star}/4\pi$.)
The first term in Eq.~(\ref{eq:steady}) represents precession of He atoms around the magnetic field; the second term represents the absorption of 1083 nm photons (``depopulation'' from the initial level); and the third term represents atoms that have absorbed 1083 nm photons, but then re-emit a photon and return to the $2^3{\rm S}^{\rm e}_1$ level (``repopulation'' into the initial level). 

\begin{widetext}

In order to accurately describe the population of the different relevant states (i.e. solve Eq.~(\ref{eq:steady})), we are only missing the repopulation term. We begin our analysis with the time evolution of the repopulation density matrix in Eq.~(51) of \cite{2017PhRvD..95h3010V} (see also Eqs.~III,10 and III,11 of \citealt{Barrat1961} for a similar derivation):
\begin{eqnarray}
\dot{\rho}_{{\rm repop},TW} & = & \int \frac{d^3\boldsymbol{k}_{\gamma}}{(2\pi)^3} \frac{d^3\boldsymbol{k}'_{\gamma}}{(2\pi)^3} \sum\limits_{\alpha,\beta,\mu,N,\nu,R} \hbar^{-4}(2\pi \hbar)^{2}\omega \omega' f(\boldsymbol{k}_{\gamma}) 
\bra{T}\boldsymbol{d}\cdot e_{\beta}^\ast(\hat{\boldsymbol k}'_{\gamma})\ket{\mu}
\bra{\mu}\boldsymbol{d}\cdot e_{\alpha}(\hat{\boldsymbol k}_{\gamma})\ket{N} 
 \nonumber \\ && 
\times\bra{\nu}\boldsymbol{d}\cdot e_{\beta}(\hat{\boldsymbol k}'_{\gamma})\ket{W} 
\bra{R}\boldsymbol{d}\cdot e_{\alpha}^\ast(\hat{\boldsymbol k}_{\gamma})\ket{\nu}
\frac{\rho_{NR} \pi \delta(\omega'-\omega + \omega_{WR})}{[(\omega - \omega_{\mu N})+i\Gamma_{2^3\rm P}/2][(\omega - \omega_{\nu R})-i\Gamma_{2^3\rm P}/2]}  + ~ \rm{h.c.} \, \textup{,} 
\label{eq:repop_0}
\end{eqnarray}
where $\boldsymbol{k}_{\gamma}$, $\boldsymbol{k}'_{\gamma}$, and $\boldsymbol{d}$ represent the incoming and outgoing photon momentum and the dipole moment operator, respectively. The $e_{\alpha (\beta)}(\hat{\boldsymbol{k}}{(')}_{\gamma})$ are the polarization unit vectors of incoming (outgoing) radiation. The indices $\alpha$ and $\beta$ run over the polarization states of the incoming and outgoing radiation, respectively. The remaining indices run over the possible initial states ($N,R$) and intermediate states ($\mu,\nu$) of the atom. The width of the intermediate states is $\Gamma_{2^3\rm P}$. The ``h.c.'' stands for Hermitian conjugate. Equation~(\ref{eq:repop_0}) is essentially the density matrix version of Fermi's Golden Rule for second-order transitions. Physically, it describes the evolution of the ground state taking into account the absorption of a 1083 nm photon ($\boldsymbol{k}_{\gamma}$) from the ground state to an excited state followed by the subsequent emission of a 1083 nm photon ($\boldsymbol{k}'_{\gamma}$). The matrix element for this process contains the dipole matrix element to excite the atom ($N\rightarrow \mu$), a propagator (denominator $\omega-\omega_{\mu N}+i\Gamma_{2^3\rm P}/2$), and the dipole matrix element for de-excitation ($\mu\rightarrow T$). The density matrix formulation has two copies of this matrix element (for $N\rightarrow T$ and $R\rightarrow W$) rather than a matrix element squared. There is an integral over the phase space for ingoing and outgoing photons ($\int d^3{\boldsymbol k}/(2\pi)^3$ and $\int d^3{\boldsymbol k}'/(2\pi)^3$, respectively), polarization sums (over $\alpha$ and $\beta$), and a phase space density for ingoing photons. The factor of $\pi \delta(\omega'-\omega+\omega_{WR})+{\rm h.c.}$ is the usual energy-conserving $\delta$-function or density of states. 

We can further simplify Eq.~(\ref{eq:repop_0}) by expanding the dot products in the matrix elements into their components and performing the sum over polarization states. For convenience when working with angular momentum coupling, we work in the spherical basis, where the basis vectors are $\hat{\boldsymbol e}_0 = \hat{\boldsymbol e}_z$ and $\hat{\boldsymbol e}_{\pm 1} = \mp\frac1{\sqrt2}(\hat{\boldsymbol e}_x \pm i\hat{\boldsymbol e}_y)$. In this basis, the dot product of two vectors can be written as ${\boldsymbol a}\cdot{\boldsymbol b} = \sum_{qs} g_{qs} a_q b_s$, where the metric tensor $g_{qs} = (-1)^q\delta_{qs}$. Note, however, that in a complex basis we must distinguish between the complex conjugate of a vector component $(a_q)^\ast$ and a component of a complex conjugate $(a^\ast)_q$. The sum over polarization states is the tensor
\begin{equation}
C_{qs}(\hat{\boldsymbol k}_{\gamma}) \equiv \sum_{\alpha = H,V} ({\boldsymbol e}_{\alpha}(\hat{\boldsymbol k}_{\gamma}))_{q} ({\boldsymbol e}^{\ast}_{\alpha}(\hat{\boldsymbol k}_{\gamma}))_{s} = 
\left(\sj{-\frac12e^{2i\phi}\sin^{2}{\theta}}{-\frac1{\sqrt2}e^{i\phi}\cos{\theta}\sin{\theta}}{-\frac12(1+\cos^2{\theta})}{-\frac1{\sqrt2}e^{i\phi}\cos{\theta}\sin{\theta}}{\sin^2{\theta}}{\frac1{\sqrt2}e^{-i\phi}\cos{\theta}\sin{\theta}}{-\frac12(1+\cos^2{\theta})}{\frac1{\sqrt2}e^{-i\phi}\cos{\theta}\sin{\theta}}{-\frac12e^{-2i\phi}\sin^{2}{\theta}} \right)
= g_{qs} - \hat k_q\hat k_s\, \textup{,}
\label{eq:polarization}
\end{equation}
where $H$ and $V$ represent the horizontal and vertical polarization states, and $q$ and $s$ are the components of the unit vectors that span the range $-1$, 0, and $+1$ (spherical basis, written in that order). A similar tensor is also defined for the outgoing radiation. With this replacement, we see that
\begin{equation}
\sum_{\alpha = H,V}
({\boldsymbol a}\cdot {\boldsymbol e}_{\alpha}(\hat{k}_{\gamma}))
({\boldsymbol b}\cdot {\boldsymbol e}_{\alpha}^\ast(\hat{k}_{\gamma}))
= \sum_{q,s} (-1)^q (-1)^s a_{q} b_{s} C_{-q,-s} = \sum_{q,s} a_{q} b_{s} [C_{qs}(\hat{\boldsymbol k}_{\gamma})]^\ast.
\end{equation}
Note the useful fact that the integral over all directions is $\int_{S^2} C_{qs}(\hat{\boldsymbol k}_{\gamma})\, d^2\hat{\boldsymbol k}_\gamma = \frac{8\pi}3g_{qs}$.

The integrals over photon momentum can be rewritten in spherical coordinates: $ \int_0^\infty dk\,\int_{S^2} d^2\hat{\boldsymbol k}\, k^{2}/(2\pi)^3$. We can simplify Eq.~(\ref{eq:repop_0}) by integrating it over incoming and outgoing photon solid angles to get
\begin{eqnarray}
\dot{\rho}_{{\rm repop},TW} & = &  \sum\limits_{\mu,N,\nu,R,q,s,n,r}\int \frac{dk_{\gamma}\,k_{\gamma}^2}{(2\pi)^3} \frac{dk'_{\gamma}\,k'_{\gamma}{^2}}{(2\pi)^3} \hbar^{-4}(2\pi \hbar)^{2} \omega \omega'\frac{8 \pi^2}{3}(-1)^{n}\delta_{n,-r}f_{\rm star}(\omega)[C_{qs}(\hat{\boldsymbol n})]^\ast \Omega_{\rm star}  \nonumber \\ 
&&\times
\bra{T}d^{n}\ket{\mu}\bra{\mu}d^{q}\ket{N} \bra{\nu}d^{r}\ket{W} \bra{R}d^{s}\ket{\nu} \frac{\rho_{NR} \delta(\omega'-\omega + \omega_{WR})}{[(\omega - \omega_{\mu N})+i\Gamma_{2^3\rm P}/2][(\omega - \omega_{\nu R})-i\Gamma_{2^3\rm P}/2]} + ~ \rm{h.c.} \, \textup{,}
\label{eq:repop_1}
\end{eqnarray}
where $C_{qs}(\hat{\boldsymbol n})$ is Eq.~(\ref{eq:polarization}) evaluated for the direction of incident starlight.

In order to further simplify Eq.~(\ref{eq:repop_1}), we need to use the assumption that the spectrum of incoming radiation is smooth and unpolarized. We convert the wavenumber integrals to frequency using $k_\gamma = \omega/c$ and $k'_\gamma = \omega'/c$. Thus we can pull the phase space density of incident photons and all smoothly varying functions of $\omega$ out of the integral, and integrate over $k'_{\gamma}$ (using the $\delta$ function) and $k_{\gamma}$ (using contour integration) to get
\begin{eqnarray}
\dot{\rho}_{{\rm repop},TW} & = &  \sum\limits_{\mu,N,\nu,R,q,s,n,r} \hbar^{-4}\frac{(2\pi \hbar)^{2}}{(2 \pi c)^{6}}\frac{8\pi^2}{3}(-1)^{n}\delta_{n,-r} [C_{qs}(\hat{\boldsymbol n})]^\ast \Omega_{\rm star} 
\bra{T}d^{n}\ket{\mu}\bra{\mu}d^{q}\ket{N} \bra{\nu}d^{r}\ket{W} \bra{R}d^{s}\ket{\nu}
\nonumber \\ 
&& \times
\frac{2\pi i\rho_{NR}f(\omega_0)\omega_0^6}{\omega_{\nu R} - \omega_{\mu N} + i\Gamma_{2^3\rm P}} 
 + ~ \rm{h.c.} \, \textup{,}
\label{eq:repop_2}
\end{eqnarray}
where again $\omega_0=1.74\times 10^{15}\,$s$^{-1}$ is the line transition frequency (used only in places where there is a negligible difference depending on which of the three lines is used).
To simplify the matrix elements appearing in Eq.~(\ref{eq:repop_2}), we will use the Wigner--Eckart theorem and the formula for double-barred matrix elements in spin-orbit coupling (see Eq.~7.1.7 of \citealt{1960amqm.book.....E}) to find
\begin{eqnarray}
D^{n}_{\mu T} &\equiv&
\langle \mu | d^n | T \rangle \nonumber \\
&=&  (-1)^{J_\mu-M_{J,\mu}+L_\mu+S_\mu+J_T + 1} \sqrt{(2J_\mu+1)(2J_T+1)} 
 \left(\tj{J_\mu}{1}{J_T}{-M_{J,\mu}}{n}{M_{J,T}}\right) \left\{\tj{L_\mu}{J_\mu}{S_\mu}{J_T}{L_T}{1}\right\} \langle n_\mu, L_\mu ||\boldsymbol{d} || n_T, L_T \rangle
 \nonumber \\
 &=&  (-1)^{M_{J,\mu} }  \sqrt{\frac{2J_\mu+1}{3}}
 \left(\tj{J_\mu}{1}{1}{-M_{J,\mu}}{n}{M_{J,T}}\right) \langle 2^3{\rm P}^{\rm o} ||\boldsymbol{d} || 2^3{\rm S}^{\rm e} \rangle\,.
\label{eq:matrix_ele.1}
\end{eqnarray}
In the second step, we used the fact that $S_\mu=S_T=1$, $L_T=0$, $L_\mu=1$, and $J_T=1$ to simplify the phase factors and $6j$ symbols.
The final double-barred matrix element acts on the orbital wave function only. Note that the matrix element going the other way is $\langle T | d^n |\mu \rangle = (-1)^n (D^{-n}_{\mu T})^\ast$ because $(d^n)^\dagger = (-1)^nd^{-n}$. We then obtain
\begin{eqnarray}
\dot{\rho}_{{\rm repop},TW} & = &  \sum\limits_{\mu,N,\nu,R,q,s,n,r} \hbar^{-4}\frac{(2\pi \hbar)^{2}}{(2 \pi c)^{6}}\frac{8\pi^2}{3}(-1)^{2n+s}\delta_{n,-r} 
[C_{qs}(\hat{\boldsymbol n})]^\ast \Omega_{\rm star}
(D^{-n}_{\mu T})^\ast D^{q}_{\mu N} D^{r}_{\nu W}(D^{-s}_{\nu R})^\ast
\frac{2 \pi i\rho_{NR}f_{\rm star}(\omega_0)\omega_0^6}{\omega_{\nu R} - \omega_{\mu N} + i\Gamma_{2^3\rm P}} \nonumber \\
 && + ~ \rm{h.c.} \, \textup{.} 
\label{eq:repop_3}
\end{eqnarray}

\subsection{Polarization of the Metastable Helium Atoms: Solution}
\label{sec:calc-sol}

Equation (\ref{eq:steady}), with Eqs.~(\ref{eq:B}), (\ref{eq:depop}), and (\ref{eq:repop_3}) describing the terms, completely specifies the density matrix evolution and can be used to solve for the steady-state solution. While we could solve for the full general density matrix, the fast precession assumption (\S\ref{sec:assumptions}) will allow us to simplify the result. The $\dot\rho_{TW}$ term can be written as
\begin{equation}
(i \omega_B (M_{J,W}- M_{J,T}) - 3A_{ul}\bar f_{1083}) \rho_{TW} + \dot\rho_{{\rm repop}, TW} = 0 ~~~\rightarrow ~~~
\rho_{TW} = \frac{\dot\rho_{{\rm repop}, TW}}{3A_{ul}\bar f_{1083} - i\omega_B (M_{J,W}- M_{J,T})}.
\end{equation}
For $T=W$, the denominator is simply $3A_{ul}\bar f_{1083}$, and by adding the three possible states of $T$, we get $1 = {\rm Tr}\rho = {\rm Tr}\,\dot\rho_{\rm repop}/(3A_{ul}\bar f_{1083})$. For $T\neq W$, the absolute value of the denominator is at least $\omega_B$ (since $M_{J,W}-M_{J,T}$ is a nonzero integer). Thus we find
\begin{equation}
|\rho_{TW}| \le \frac{{\rm Tr}\,\dot\rho_{\rm repop}}{\omega_B}
= \frac{3A_{ul}\bar f_{1083}}{\omega_B} = \frac{B_{\rm I}}{B}\ll 1
~~~{\rm for}~~~T\neq W ~~{\rm and}~~ B\gg B_{\rm I}.
\end{equation}
That is, in our ``fast precession'' approximation, the off-diagonal terms in the density matrix are small compared to 1, and we drop them in what follows.

We can further simplify Eq.~(\ref{eq:repop_3}) by expanding the matrix elements using Eq.~(\ref{eq:matrix_ele.1}) and utilizing the assumptions stated in \S\ref{sec:assumptions} (mainly fast precession) in order to obtain
\begin{eqnarray}
\dot{\rho}_{{\rm repop},TT} & = & \sum_{J_\mu=0}^2 \sum_{M_{J,\mu}=-J_\mu}^{J_\mu} \sum_{M_{J,N}=-1}^1
 \frac{\Omega_{\rm star} \omega_{0}^6f_{\rm star}(\omega_0)\rho_{NN}}{3\pi \hbar^2 c^6 \Gamma_{2^3\rm P}}(-1)^{{M_{J,N}}-{M_{J,\mu}}}[C_{{M_{J,\mu}}-{M_{J,N}},{M_{J,N}}-{M_{J,\mu}}}(\hat{\boldsymbol n})]^\ast \frac{(2 J_\mu+1)^2}{9} \nonumber \\
&& \times \left|\langle 2^3{\rm P}^{\rm o} ||\boldsymbol{d} || 2^3{\rm S}^{\rm e} \rangle \right|^4
\left(\tj{J_\mu}{1}{1}{-M_{J,\mu}}{M_{J,\mu}-M_{J,T}}{M_{J,T}}\right)^2\left(\tj{J_\mu}{1}{1}{-M_{J,\mu}}{M_{J,\mu}-M_{J,N}}{M_{J,N}}\right)^2 + \rm{h.c.} \, \textup{.}
\label{eq:repop_4}
\end{eqnarray}
We can also work in the approximation that $\Gamma_{2^3\rm P} \approx A_{2^3P^o\rightarrow 2^3S^e}$ because we are assuming that the atoms are far enough from the star that the incident radiation is not strong enough to stimulate emission of radiation (mean occupation number $\ll 1$, so spontaneous emission dominates). Simplifying Eq.~(\ref{eq:repop_4}) with this approximation and using the dipole emission formula
\begin{equation}
A_{\rm 2^3P^o\rightarrow 2^3S^e} =
\frac{4\omega_0^3}{9\hbar c^3}|\langle 2^3P^o||{\boldsymbol d}||2^3S^e\rangle|^2 ,
\end{equation}
and using appropriate trigonometric identities, we get
\begin{eqnarray}
\label{eq:repop_dot}
\dot{\rho}_{{\rm repop},TT} & = & \frac{3\tilde{A}}{4\pi} M_{{\rm repop},{TN}}\, \rho_{NN} \, \textup{,} 
\end{eqnarray}
where we introduce
\begin{equation}
    \label{eq:constant}
    \tilde{A} \equiv \Omega_{\rm star} f_{\rm star}(\omega_0)A_{2^3P^o\rightarrow 2^3S^e} \, ~~ \textup{and} ~~
    {\bf M}_{\rm repop} = \frac{1}{144}\left(\sj{93+7\cos2\theta_\star}{35+\cos2\theta_\star}{21+7\cos2\theta_\star}{30-14\cos2\theta_\star}{74-2\cos2\theta_\star}{30-14\cos2\theta_\star}{21+7\cos2\theta_\star}{35+\cos2\theta_\star}{93+7\cos2\theta_\star} \right) \, \textup{.}
\end{equation}
For the diagonal elements of the density matrix, the first term in Eq.~(\ref{eq:steady}) will go to zero. This simplifies  Eq.~(\ref{eq:steady}) to
\begin{eqnarray}
\label{eq:steady2}
\left({\bf M}_{\rm repop}- \mathds{1} \right)\rho_{TT} = 0 \, \textup{.}
\end{eqnarray}
Physically, ${\bf M}_{\rm repop}$ is a matrix of transition probabilities from one state to another, so ${\bf M}_{\rm repop}$ has a single eigenvalue that is equal to 1 because we care about steady-state solutions. This means all solutions to Eq.~(\ref{eq:steady2}) are proportional to the corresponding eigenvector. We can set the normalization by requiring ${\rm Tr}\,\rho=1$, that is, total probability unity. This gives the components of $\rho_{TT}$ as 
\begin{eqnarray}
\label{eq:rho_vec}
\rho_{TT} & = \left( \frac{\displaystyle 35+\cos{2\theta_\star}}{\displaystyle 100-12\cos{2\theta_\star}}, \frac{\displaystyle 15-7\cos{2\theta_\star}}{ \displaystyle 50-6\cos{2\theta_\star}}, \frac{\displaystyle 35+\cos{2\theta_\star}}{ \displaystyle 100-12\cos{2\theta_\star}} \right).
\end{eqnarray}
\end{widetext}
The left panel of \autoref{fig:rho_plot} shows the population of atoms in the $2^3{\rm S}^{\rm e}_1$ state from the different states within 2$^3$P$^{\rm o}_{0,1,2}$ as a function of incoming radiation.

\begin{figure*}
    \centering
    \includegraphics[width=0.495 \textwidth]{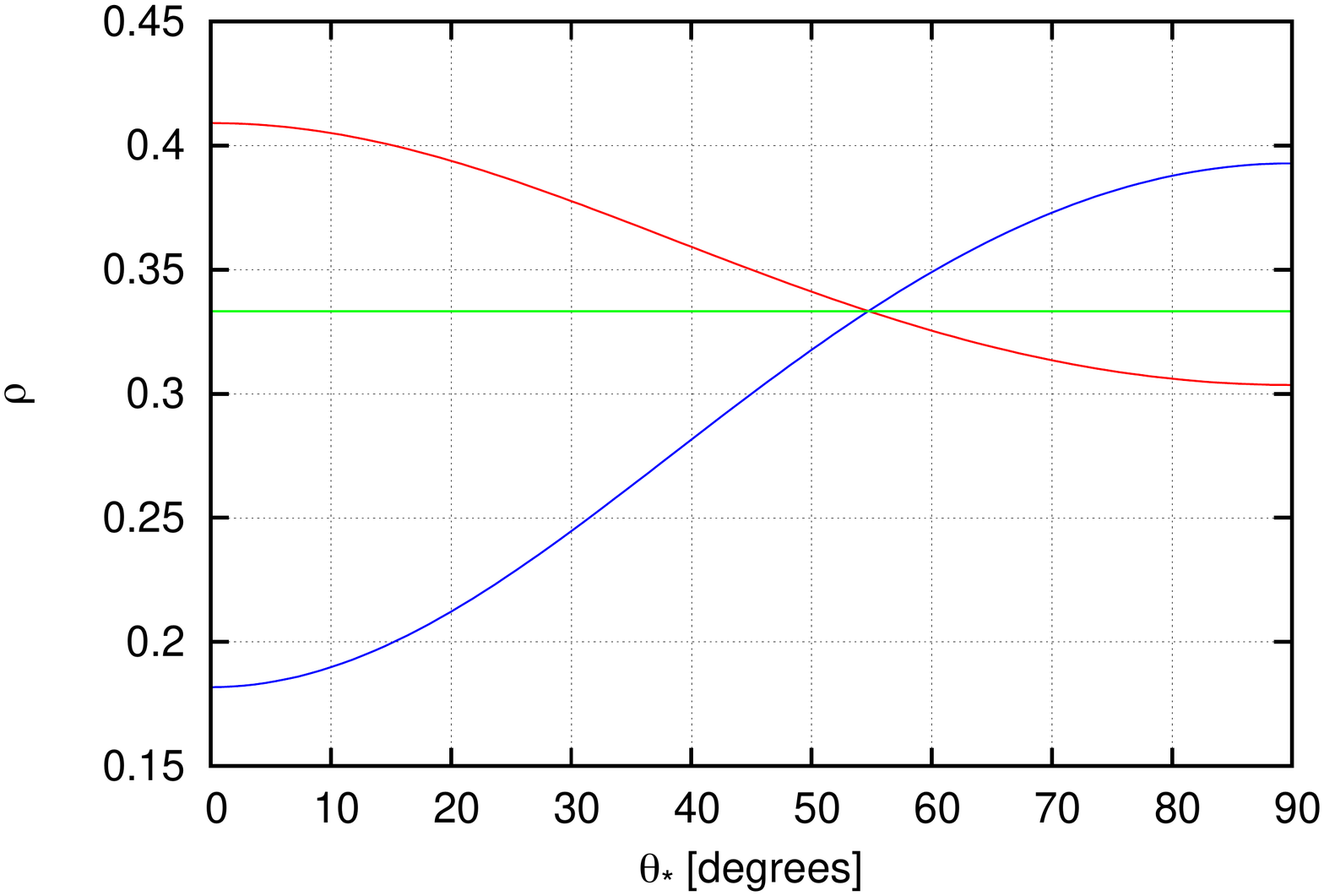}
    \includegraphics[width=0.495 \textwidth]{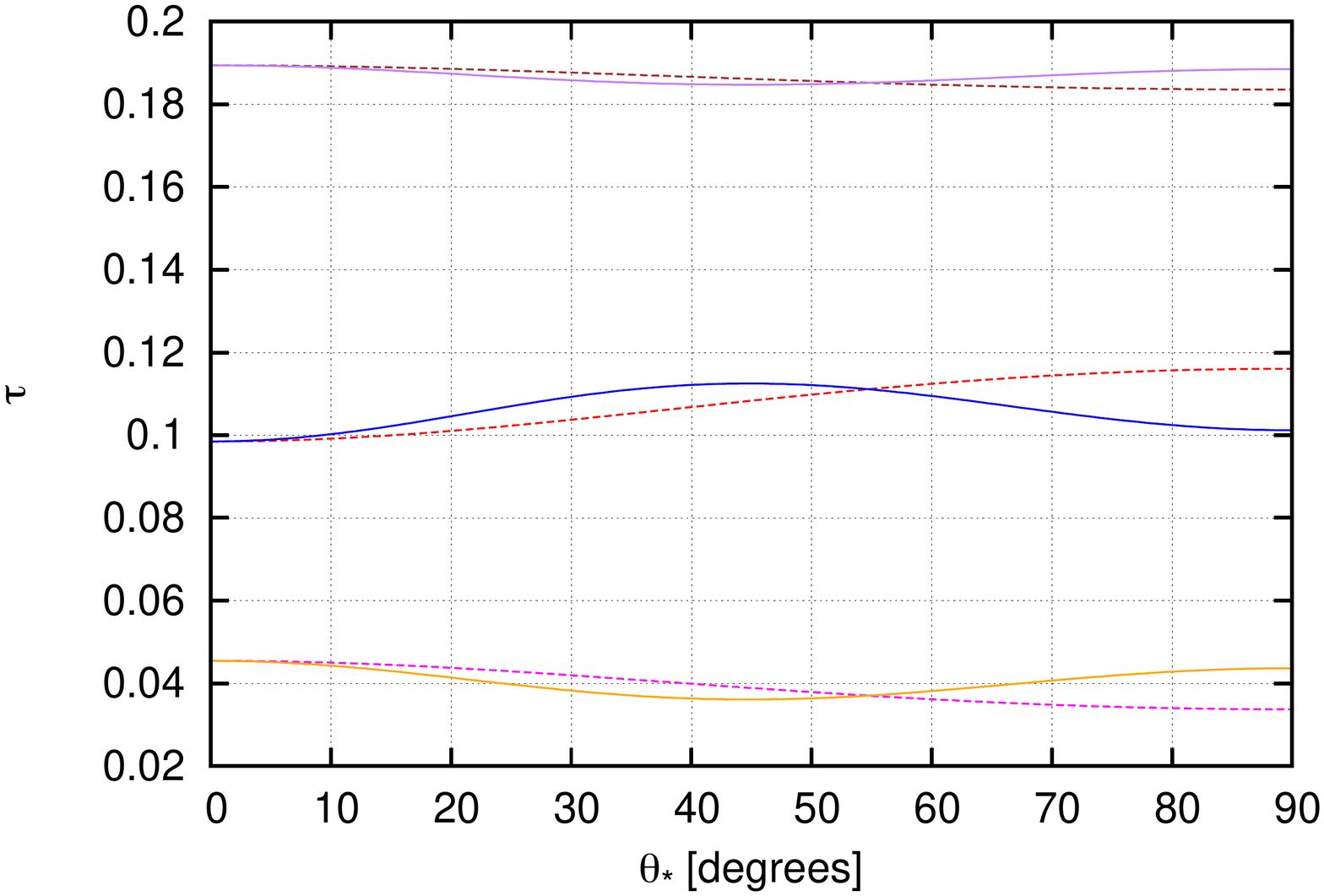}
    \caption{Left panel: the blue curve is a plot of $\rho_{-1-1}$ and $\rho_{11}$, and red is the $\rho_{00}$ as a function of the direction of incident radiation. The green constant line is the symmetric average of each component, that is, $\frac13$. Right panel: this is a plot of the $\tau^{J_{f}}_{\left(H,V\right)}$ as a function of incident radiation. The lowest two curves (magenta and orange) are $\tau^{0}_{\left(H,V\right)}$, the middle curves (red and blue) are $\tau^{1}_{\left(H,V\right)}$, and the top two curves (brown and purple) are $\tau^{2}_{\left(H,V\right)}$ (we use the solid line for the vertical polarization and the dashed line for horizontal polarization).}
    \label{fig:rho_plot}
\end{figure*}

\subsection{Polarization-selective Absorption in the 1083 nm Triplet}
\label{sec:analytic_results}

Finally, we have the required ingredients to compute the absorption optical depth for horizontally -- and vertically -- polarized light. The optical depth at line center is given by
\begin{eqnarray}
\label{eq:tau_expansion1}
\tau^{J_{f}}_{\left(H,V\right)} &\propto& \sum\limits_{{M_{J_{f}}},M_{J}} |\langle 2^3{\rm P}^{\rm o}_{J_{f}},M_{J_{f}} |\boldsymbol{d} \cdot \boldsymbol e_{\left(H,V\right)}| 2^3{\rm S}^{\rm e},M_{J} \rangle|^2
\nonumber \\ && \times \rho_{M_{J}M_{J}},
\end{eqnarray}
where $J_{f}$ is the total angular momentum of the final state (0, 1, or 2), and $e_{\left(H,V\right)}$ are the unit polarization vectors for horizontal and vertical light, respectively. The normalization factor depends on the metastable helium column density and velocity dispersion. The functional forms of Eq.~(\ref{eq:tau_expansion1}) as a function of $\theta_\star$ are
\cmnt{\begin{eqnarray}
    \label{eq:tau_0_H}
     \tau^{0}_{H} & \propto & \frac{35+\cos{2\theta_\star}}{36(25-3\cos{2\theta_\star})} \, ,\\
    \label{eq:tau_0_V}
    \tau^{0}_{V} & \propto & \frac{(35+\cos{2\theta_\star})\cos^2{\theta_\star}}{36(25-3\cos{2\theta_\star})} + \frac{(30-14\cos{2\theta_\star})\sin^2{\theta_\star}}{36(25-3\cos{2\theta_\star})} \, ,\\
    \label{eq:tau_1_H}
    \tau^{1}_{H} & \propto & \frac{13\left(\cos{2\theta_\star}-5\right)}{72\cos{2\theta_\star}-600} \, , \\
    \label{eq:tau_1_V}
    \tau^{1}_{V} & \propto & \frac{(30-14\cos{2\theta_\star})\cos^2{\theta_\star}}{24(25-3\cos{2\theta_\star})} + \frac{(35+\cos{2\theta_\star})\sin^2{\theta_\star}}{12(25-3\cos{2\theta_\star})} + \frac{(35+\cos{2\theta_\star})\cos^2{\theta_\star}}{24(25-3\cos{2\theta_\star})} \, , \\
    \label{eq:tau_2_H}
    \tau^{2}_{H} & \propto & \frac{5\left(7\cos{2\theta_\star}-67\right)}{72(3\cos{2\theta_\star}-25)} \, , ~~{\rm and}\\
    \label{eq:tau_2_V}
    \tau^{2}_{V} & \propto & \frac{(30-14\cos{2\theta_\star})\cos^2{\theta_\star}}{24(25-3\cos{2\theta_\star})} + \frac{7(35+\cos{2\theta_\star})\cos^2{\theta_\star}}{72(25-3\cos{2\theta_\star})} + \frac{(35+\cos{2\theta_\star})\sin^2{\theta_\star}}{12(25-3\cos{2\theta_\star})} + \frac{(30-14\cos{2\theta_\star})\sin^2{\theta_\star}}{18(25-3\cos{2\theta_\star})} \, .
\end{eqnarray}}
\begin{eqnarray}
    \label{eq:tau_0_H}
     \tau^{0}_{H} & \propto & \frac{35+\cos{2\theta_\star}}{36(25-3\cos{2\theta_\star})} \, ,\\
    \label{eq:tau_0_V}
    \tau^{0}_{V} & \propto & \frac{65-8\cos2\theta_\star+15\cos^22\theta_\star}{72(25-3\cos{2\theta_\star})} \, ,\\
    \label{eq:tau_1_H}
    \tau^{1}_{H} & \propto & \frac{13\left(5-\cos{2\theta_\star}\right)}{24(25-3\cos{2\theta_\star})} \, , \\
    \label{eq:tau_1_V}
    \tau^{1}_{V} & \propto & \frac{ 135-16\cos 2\theta_\star-15\cos^22\theta_\star }{48(25-3\cos{2\theta_\star})} \, , \\
    \label{eq:tau_2_H}
    \tau^{2}_{H} & \propto & \frac{5\left(67-7\cos{2\theta_\star}\right)}{72(25-3\cos{2\theta_\star})} \, , ~~{\rm and}\\
    \label{eq:tau_2_V}
    \tau^{2}_{V} & \propto & \frac{5(133 -16\cos 2\theta_\star + 3\cos^2 2\theta_\star)}{144(25-3\cos{2\theta_\star})} \, .
\end{eqnarray}
The right panel of Figure \ref{fig:rho_plot} shows the plot of $\tau^{J_{f}}_{\left(H,V\right)}$ for each $J_{f}$ and horizontal and vertical polarization. We see that for the same $J_{f}$, the values of $\tau^{J_{f}}_{\left(H,V\right)}$ are nearly the same, and there are never any crossings of $\tau^{J_{f}}_{\left(H,V\right)}$ for different $J_{f}$.


We can now compute a predicted optical depth in each of the two polarizations. We assign each line a Gaussian profile, given by some velocity width $\Delta v$. The normalization factors are all the same; they depend on the total column density of metastable helium atoms, which is not computed in this section. It could be taken by fitting to observations (here we set the peak optical depth to 0.05) or from a theory calculation similar to \citet{OklopcicHirata2018}. The result is
\begin{equation}
    \label{eq:tau_expansion}
\tau\left(\lambda\right)_{\left(H,V\right)} = \sum\limits_{J_{f}}Y\tau^{J_{f}}_{\left(H,V\right)}\exp{-\frac{\left(\lambda - \lambda_{J_{f}}\right)^2}{\Delta \lambda^2}}
     \, \textup{,}
\end{equation}
where $\Delta\lambda=(\Delta v/c)\lambda_0$, $\Delta v$ is the velocity width,\footnote{We define this to be $\sqrt2$ times the standard deviation of the \\ line-of-sight velocity; see Eq.~(5.43) of \\ \citet{Landi2004}. This is consistent with \\ the input parameter to {\sc Hazel}.} $\lambda_{0}$ is $1083$ nm, and $Y$ is the normalization set by observations.

\begin{figure*}
    \centering
    \includegraphics[width=0.45 \textwidth]{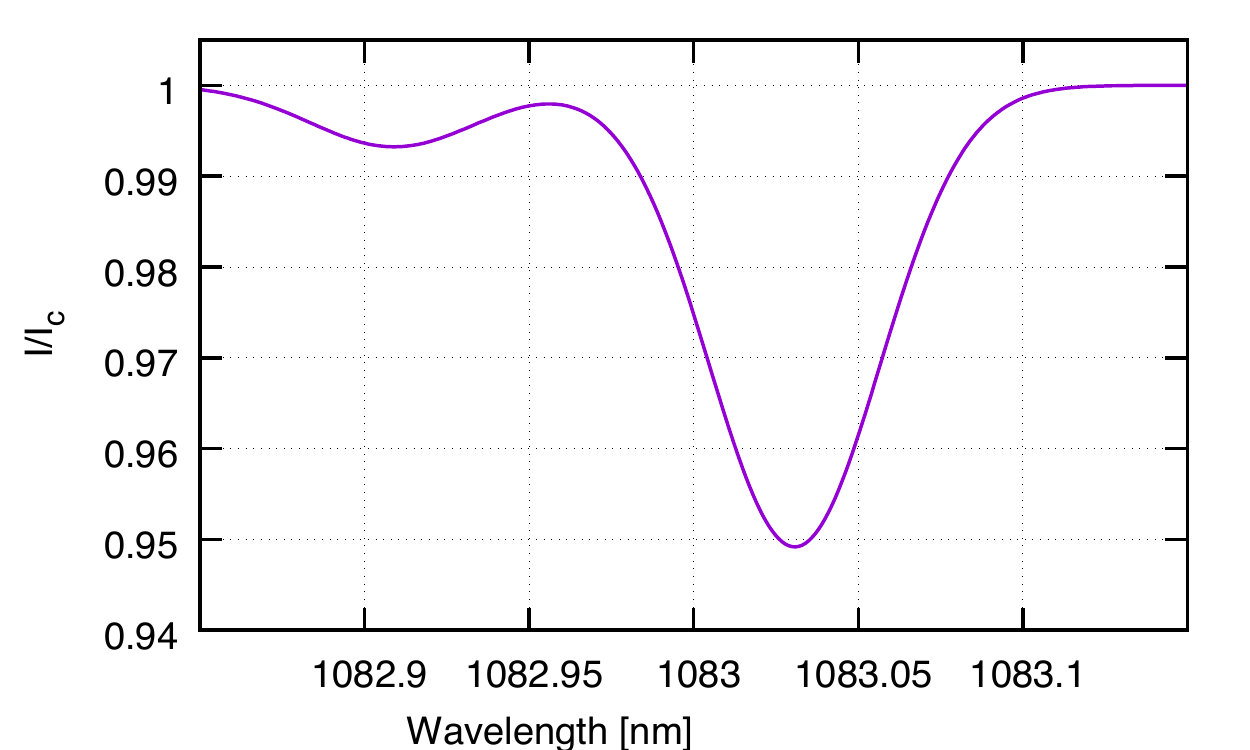}
    \includegraphics[width=0.45
    \textwidth]{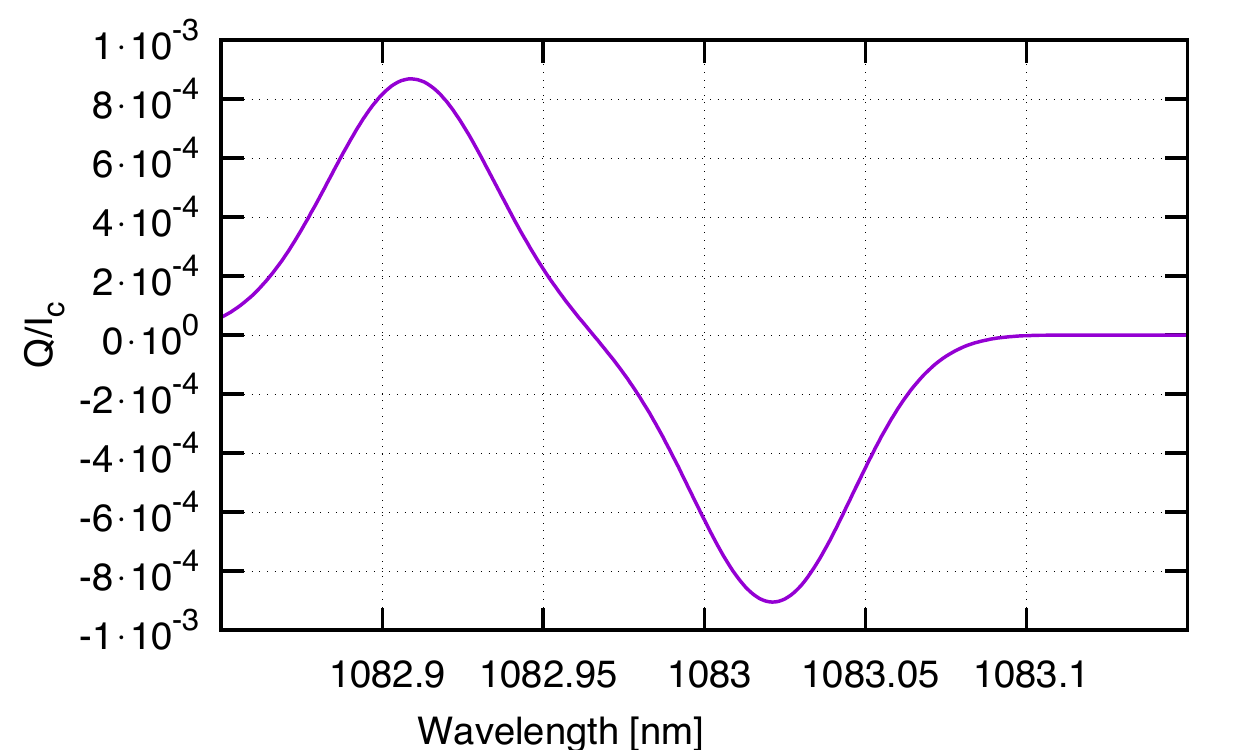}
    \caption{ Left panel: analytic solution for total intensity ($I/I_{c}$). Right panel: the analytic solution for the linear polarization Stokes parameter $Q/I_c$ for a magnetic field perpendicular to the line of sight, $\theta_\star = \frac\pi2$.}
    \label{fig:analytic_plot}
\end{figure*}

The total intensity and Stokes parameter $Q$ are plotted as functions of wavelength in Figure \ref{fig:analytic_plot}. In the left panel of Figure \ref{fig:analytic_plot}, we see the absorption features in the spectrum relative to the out-of-transit signal ($I_{c}$). The weaker and bluer absorption feature is due to the transition from 2$^3$S to 2$^3$P$_{J=0}$. The more prominent absorption feature is due to the transitions from 2$^3$S to 2$^3$P$_{J=1,2}$. Since the $J=1$ and $J=2$ states are very close in energy, their absorption signals are blended; their combined oscillator strengths are a factor of 8 larger than the $J=0$ transition, hence the deeper absorption feature. For the right panel in Fig.~\ref{fig:analytic_plot}, the Stokes parameter $Q$ undergoes a continuous transition from positive (polarization perpendicular to the $B$-field) in the blue feature to negative (polarization parallel to the $B$-field) in the red feature. This positive--negative behavior had to happen, because in an $^3$S$^{\rm e}\rightarrow\,^3$P$^{\rm o}$ absorption the spin degree of freedom does not participate in the transition and the initial orbital state is isotropic; thus by sum rules the frequency-integrated cross section $\int \sigma(\nu)\, d\nu$ is the same for both incident photon polarizations regardless of how the atom is spin-polarized. This means that if one line has linear polarization, the other must have the opposite linear polarization, giving the shape of the curve in the right panel of \autoref{fig:analytic_plot}.

\section{Numerical Calculation}
\label{sec:numerical}

We now turn to numerical calculations. We describe our use of the \textsc{Hazel} code in \S\ref{ss:hazel}, and then we proceed to consider a range of cases with a uniform magnetic field (including several orientations and strengths). We defer discussion of nonuniform fields and resulting suppression of the polarization signal to \S\ref{ss:twocomponent}.

\subsection{\textsc{Hazel} Code}
\label{ss:hazel}

We use the publicly available code \textsc{Hazel} (version 2.0),\footnote{\url{ https://github.com/aasensio/hazel2}} developed by \citet{AsensioRamos2008}, to calculate the effect of a planetary magnetic field on the observed polarization of the helium 1083~nm line. \textsc{Hazel} (\textit{HAnle and ZEeman Light}) calculates the Stokes profiles of radiation passing through a constant-property slab of helium in the presence of a magnetic field\footnote{\textsc{Hazel} can also be used in `inversion mode' to infer model parameters from a set of observed Stokes parameters.}. The code takes into account all the relevant physical processes: optical pumping; atomic-level polarization; level crossing and repulsion; and the Hanle, Zeeman, and Paschen-Back effects. The helium model used in \textsc{Hazel} includes transitions between the following atomic levels: $2s\,^3S$, $3s\,^3S$, $2p\,^3P$, $3s\,^3P$, and $3d\,^3D$.

A constant-property slab of neutral helium is assumed to be located at a distance $h$ from the light source whose incident spectrum is modeled after the Sun. The slab's optical depth in the red component of the helium line ($\tau$) and the line width ($\Delta v$) are free parameters that determine the shape of the Stokes $I$ profile. For our fiducial case, we choose the values of parameters that produce an absorption line similar to those observed in close-in exoplanets \citep[e.g.][]{Allart2018,Allart2019, Nortmann2018}: $h\sim 0.05$~au, $\tau = 0.05$, and $\Delta v = 10.0$~km~s$^{-1}$.

\begin{figure}
    \centering
    \includegraphics[width=0.45\textwidth]{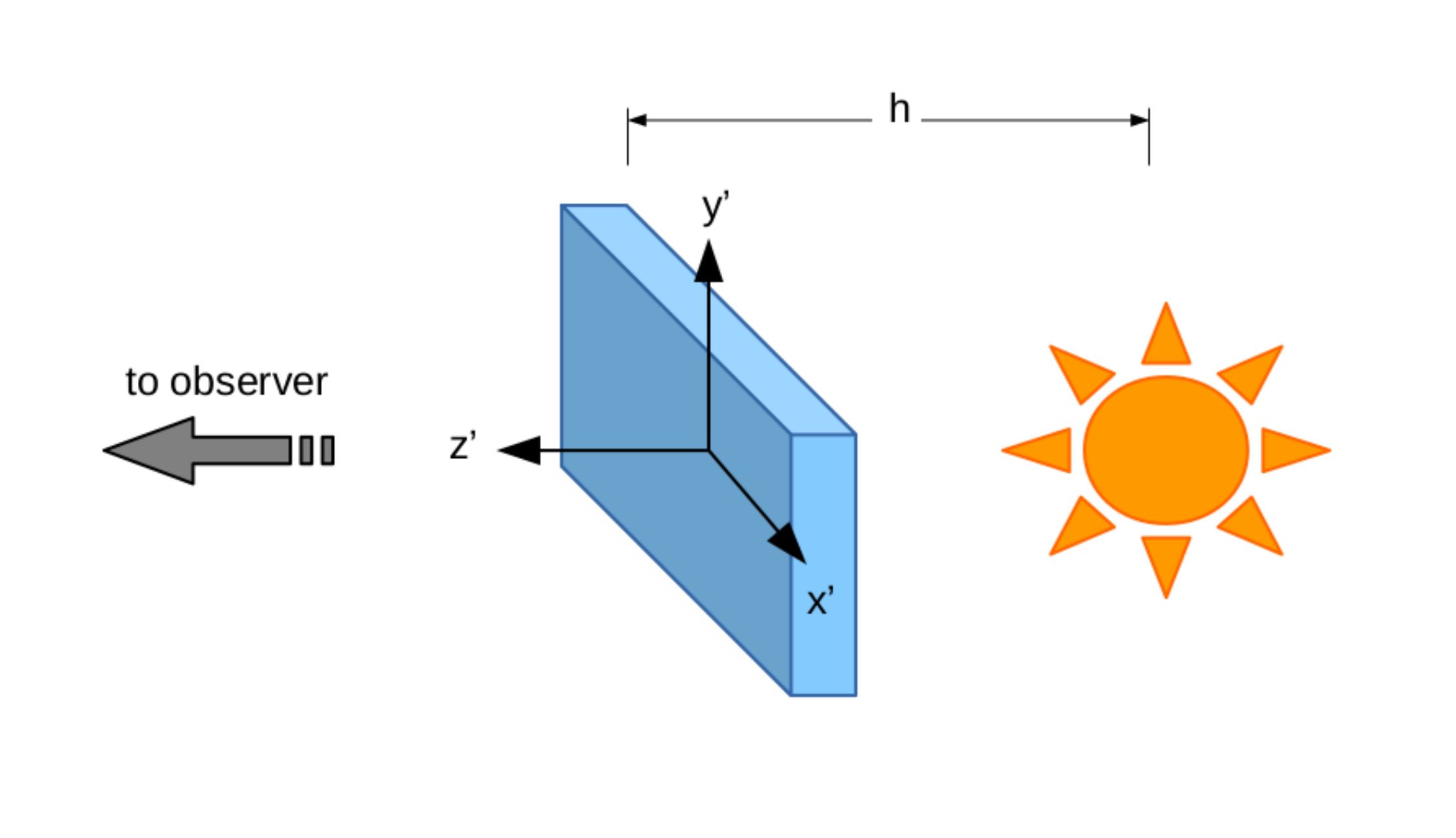}
    \caption{Problem setup for the \textsc{Hazel} code used in our analysis: a constant-property slab of neutral helium is placed at a distance $h\sim 0.05$~au from a star with spectral properties of the Sun, along the $z\prime$--axis which connects the star and the observer. A magnetic field characterized by ($B_{x^\prime}$, $B_{y^\prime}$, $B_{z^\prime}$) permeates the slab.}
    \label{fig:geometry_hazel}
\end{figure}

Our choice of the problem geometry is shown in \autoref{fig:geometry_hazel} (note that the coordinate system defined here is different from the one introduced in \S\ref{ss:specpol}): the slab is located on the $z^\prime$-axis, that is, the line of sight from the observer to the star. The magnetic field strengths along all three components of the coordinate system ($B_{x^\prime}$, $B_{y^\prime}$, $B_{z^\prime}$) are free parameters of the model. In the following sections, the magnetic field component parallel to the line of sight is denoted by $B_\parallel = B_{z^\prime}$, and the component perpendicular to the line of sight is $B_\perp = \sqrt{B_{x^\prime}^2 + B_{y^\prime}^2}$.

\subsection{Polarization in the Presence of a Longitudinal Magnetic Field}

First we investigate the dependence of radiation polarization on the presence of a magnetic field along the line of sight. This component of the magnetic field induces circular polarization because of the longitudinal Zeeman effect: in an external magnetic field, atomic levels with total angular momentum $J$ split into $(2J+1)$ sublevels, with the splitting proportional to the magnetic field strength. The wavelength shifts between different components of the spectral line result in line polarization.

\begin{figure*}
    \centering
    \includegraphics[width=0.95\textwidth]{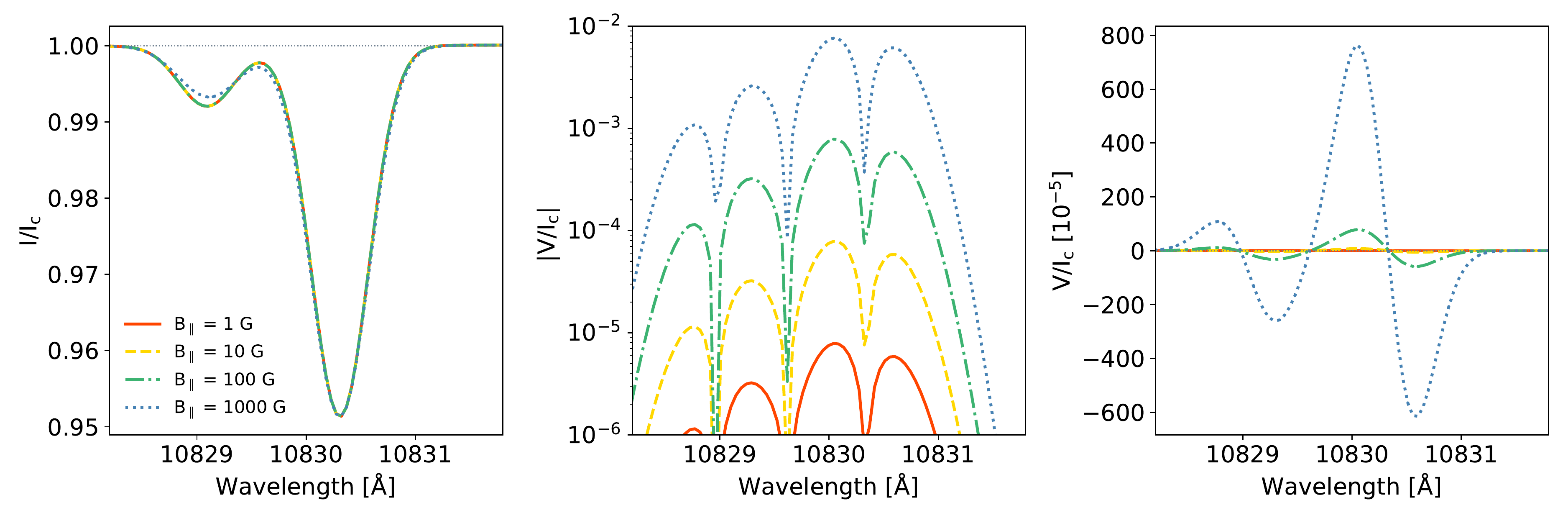}
    \caption{Magnetic field along the line of sight produces circular polarization in the helium 1083~nm line. The left panel shows the radiation intensity (Stokes $I$). The absorption line profile resembles those observed in transiting exoplanets. The middle panel shows the magnitude of the circular polarization signal (Stokes $V$). The full Stokes $V$ profile is shown in the right panel.}
    \label{fig:hazel_results1}
\end{figure*}

\autoref{fig:hazel_results1} shows the radiation intensity (Stokes $I$) and circular polarization (Stokes $V$) calculated using the \textsc{Hazel} code with the setup described in the previous section. We vary the strength of the magnetic field in the $z^\prime$ direction, while keeping all other parameters fixed. The middle panel shows the amplitude of the Stokes $V$ parameter on a logarithmic scale, ranging from $\sim {\rm few}\times 10^{-6}$ for $B_\parallel = 1$~G to $\sim {\rm few}\times 10^{-3}$ for $B_\parallel = 1$~kG. The right panel shows the Stokes $V$ line profile on a linear scale, consisting of a positive and a negative part in both the blue and red components of the helium 1083~nm line.

\subsection{Polarization in the Presence of a Transverse Magnetic Field}

\begin{figure*}
    \centering
    \includegraphics[width=0.9\textwidth]{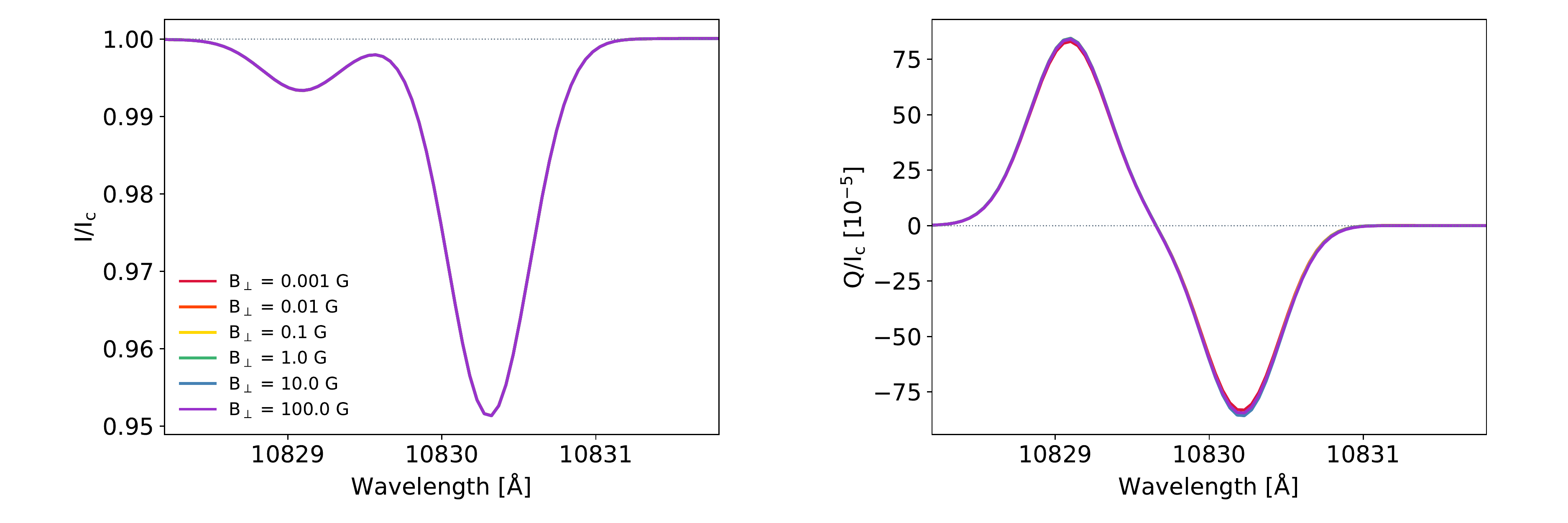}
    \caption{Transverse magnetic field produces a linear polarization signal that is independent of the magnetic field strength over many orders of magnitude. The left panel shows the radiation intensity, and the right panel shows the induced linear polarization (Stokes $Q$). In the optically thin case, the red and the blue components of the helium 1083~nm triplet have the same magnitude of linear polarization, with opposite signs.}
    \label{fig:hazel_results2}
\end{figure*}

Next we investigate the radiation polarization signal in the presence of a magnetic field perpendicular to the line of sight. This component of the magnetic field modifies the atomic-level polarization induced by anisotropic radiation and creates linear polarization in the helium 1083~nm line through the Hanle effect \citep{TrujilloBueno2002}. \autoref{fig:hazel_results2} shows the linear polarization signal (Stokes $Q$) in the presence of $B_\perp$ of strength ranging from 0.001~G to 1~kG. Linear polarization is induced by the presence of a magnetic field perpendicular to the line of sight, but the polarization signal does not depend on the magnetic field strength in the analyzed range because it is in the `saturated Hanle regime.' The amplitude of the linear polarization signal is on the order of $10^{-4}-10^{-3}$, and the blue and red components of the helium 1083~nm line have opposite polarities of the same magnitude (in the optically thin case; see \S\ref{sec:optical_depth} for more details on the optically thick case). 
Note the excellent agreement between the \textsc{Hazel} calculation in \autoref{fig:hazel_results2} and the analytic calculation in \autoref{fig:analytic_plot}.

\subsection{Polarization in the Presence of a Magnetic Field of Arbitrary Orientation}

\begin{figure*}
    \centering
    \includegraphics[width=0.95\textwidth]{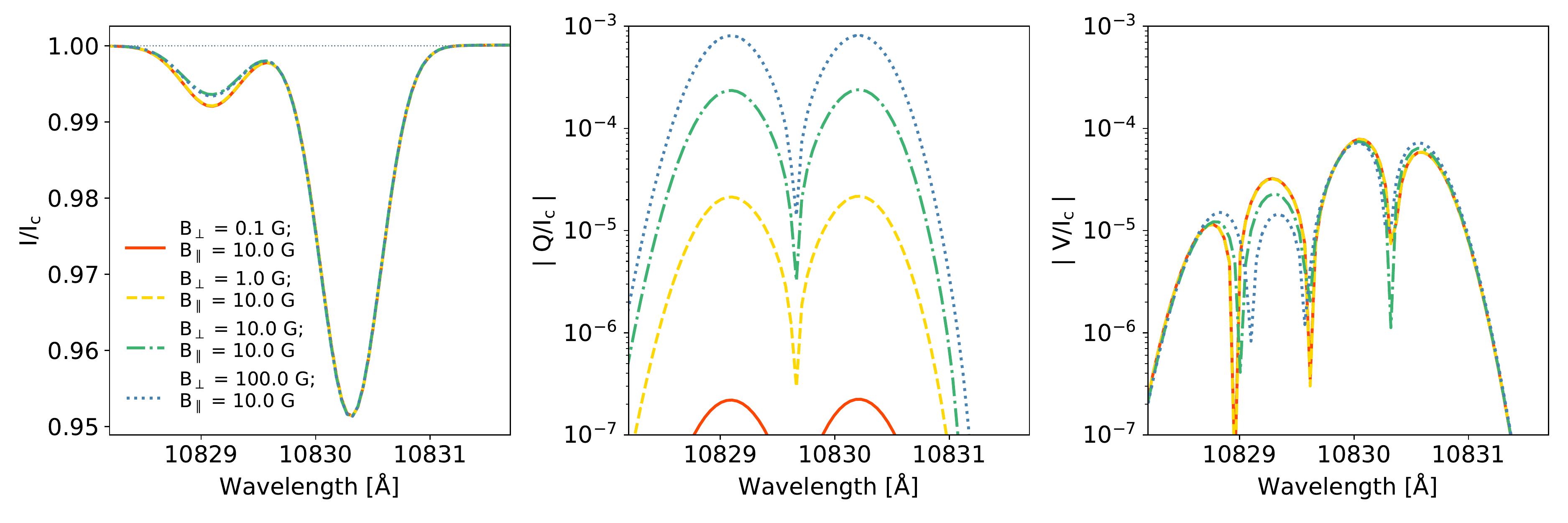}
    \caption{Magnetic field component along the line of sight ($B_\parallel$) breaks the degeneracy in the linear polarization signal induced by the perpendicular component ($B_\perp$). }
    \label{fig:hazel_results3}
\end{figure*}

In the most general case, we consider a magnetic field with nonzero components in both the line-of-sight and perpendicular directions. We consider a range of $B_\perp$ between 0.1 G and 100 G. The addition of a $B_\parallel$ (in this case $B_\parallel=10$~G) breaks the degeneracy in the linear polarization signal and induces a circular polarization signal (see \autoref{fig:hazel_results3}). For $B_\perp \ll B_\parallel$, the linear polarization is significantly reduced compared to the $B_\parallel = 0$ case shown in \autoref{fig:hazel_results2}. For $B_\perp \gg B_\parallel$, the linear polarization signal remains virtually the same as in the $B_\parallel = 0$ case. This makes sense in the context of the analytic model (\S\ref{sec:analytic}), because once the field is strong enough so that the metastable helium atoms precess many times between interactions ($B\gg B_{\rm I}$), it is the direction rather than strength of the field that determines the mean polarization of the atoms. For $B_\perp\ll B_\parallel$, we are looking down the field ($\theta_\star\approx 0$).

On the other hand, the presence of a $B_\perp$ field has a less dramatic effect on the circular polarization signal, whose amplitude remains at the same level as in the $B_\perp=0$ case, shown in \autoref{fig:hazel_results1}. This is because, while the Zeeman splitting is proportional to the total field $B$, the fractional circular polarization of each component (for $\Delta M_J=\pm 1$, i.e., the components that have circular polarization) is proportional to $\cos\theta_\star$; thus, when the Zeeman splitting is small compared to the line width ($B\ll B_{\rm IV}$), the amplitude of the negative--positive pattern in Stokes $V$ is proportional to $B\cos\theta_\star = B_\parallel$ \citep{1913ApJ....38...99S}. However, for $B_\perp \gg B_\parallel$, the line profile changes so that the positive--$V$ and the negative--$V$ parts of both the blue and red components of the helium 1083~nm multiplet have roughly equal amplitudes.

\subsection{Dependence on Optical Depth}
\label{sec:optical_depth}

\begin{figure*}
    \centering
    \includegraphics[width=0.95\textwidth]{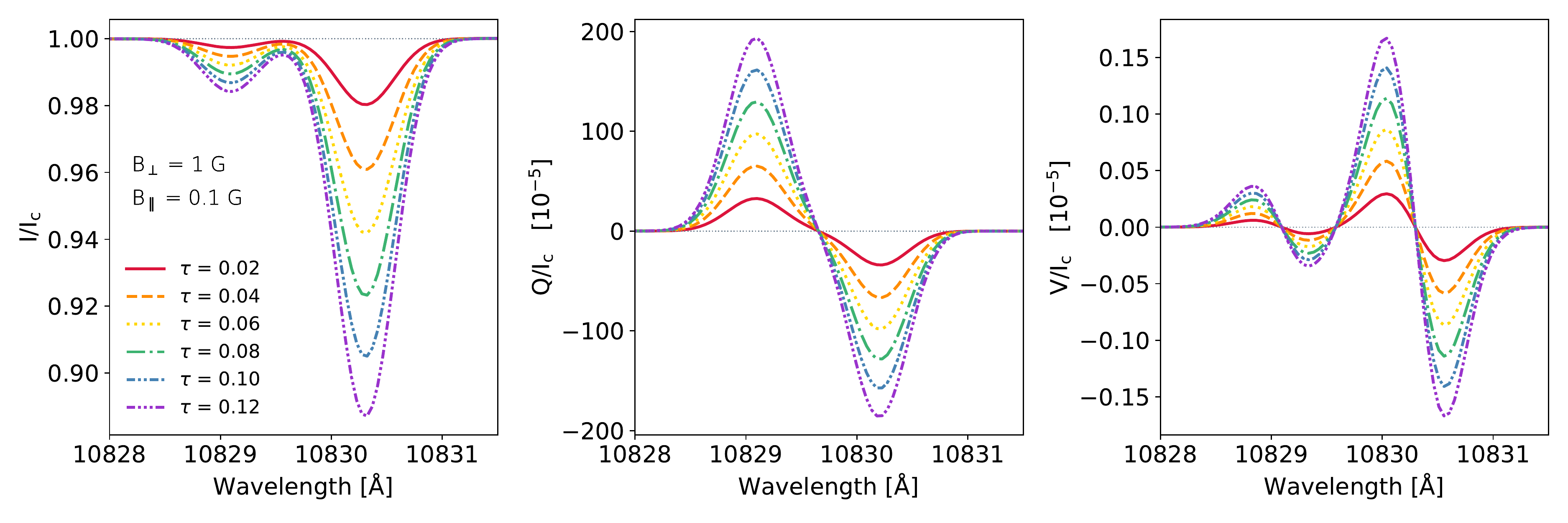}
    \caption{Radiation intensity (left), linear polarization (middle), and circular polarization (right) for different values of optical depth, with all other parameters fixed, including the magnetic field strength and orientation.
    }
    \label{fig:hazel_results4}
\end{figure*}

In the previous examples, we kept the optical depth in the red component of the helium 1083~nm line fixed at $\tau =0.05$. In \autoref{fig:hazel_results4} we show how the polarization signal changes with changing the optical depth of the medium. In the optically thin regime, the polarization signal increases linearly with optical depth, with roughly equal polarization amplitudes in the blue and red components. As optical depth grows ($\tau \gtrsim 0.1$), the linear polarization amplitude in the red component becomes smaller than the linear polarization of the blue component.

\begin{figure*}
    \centering
    \includegraphics[width=0.9\textwidth]{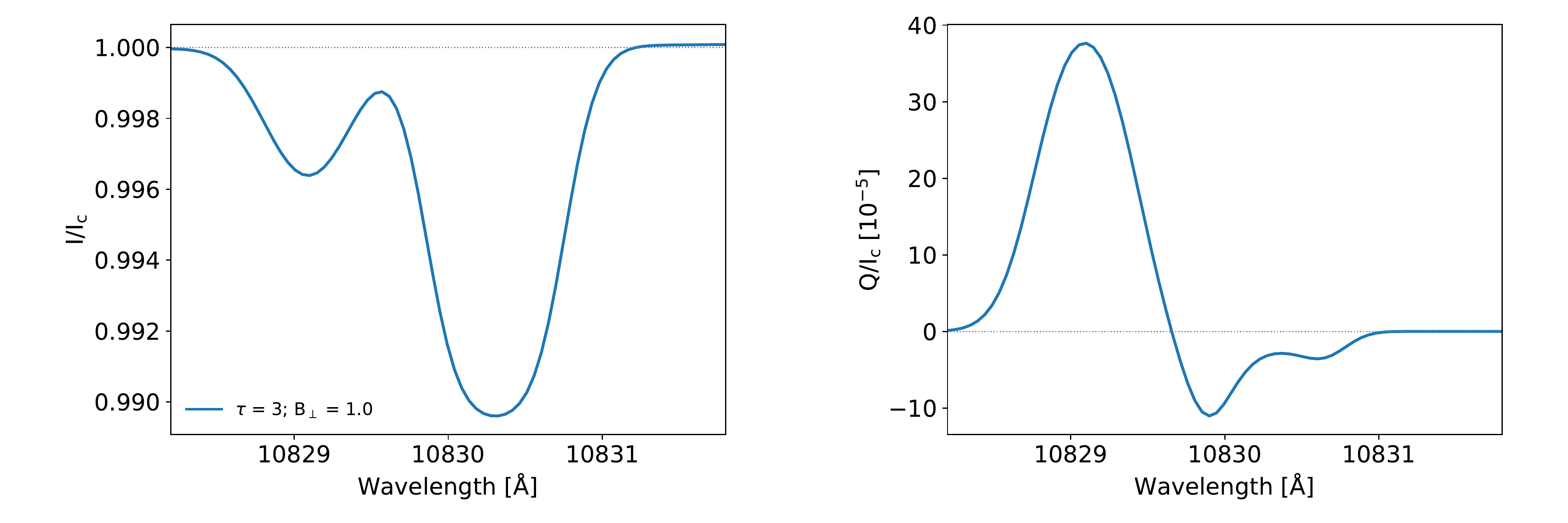}
    \caption{A slab of helium which is optically thick in the red component of the 1083 nm line produces a radiation intensity profile with a smaller red-to-blue line ratio compared to the optically thin case. For $\tau =3$, we obtain the line ratio $\sim 3:1$ (left panel), similar to that observed in HD~189733b \citep{Salz2018}. The right panel shows the reduced linear polarization in the optically thick red component of the helium line.}
    \label{fig:hazel_results5}
\end{figure*}

\autoref{fig:hazel_results5} shows the results for $\tau =3$ in the red component of the line. This example is motivated by observations of the exoplanet HD~189733b \citep{Salz2018}, whose line profile has a low red-to-blue component ratio of $\sim 3:1$ (the optically thin ratio is $\sim 8:1$). The linear polarization signal in the optically thick red component gets significantly reduced in amplitude and its profile modified compared to the blue component, which is still in the optically thin regime.

\section{Discussion} \label{sec:discussion}

In this section, we examine the prospects for observing the calculated polarization at 1083 nm in transits of close-in exoplanets and discuss potential challenges for observations.

\subsection{Combining Polarization Signals from Different Parts of the Atmosphere}
\label{ss:twocomponent}

Our results presented in the previous section were obtained for homogeneous slabs of helium permeated with uniform magnetic fields. In real observations, the signal would consist of contributions from different parts of the exoplanet atmosphere, characterized by different physical properties, including different geometries of the magnetic field. Magnetic fields in the atmospheres of highly irradiated, close-in exoplanets can have complex geometries; toroidal magnetic fields can be induced in the atmosphere by winds of ionized gas moving through the global, deep-seated, poloidal magnetic field \citep[e.g.][]{RogersKomacek2014, RogersMcElwaine2017}. Here we investigate the circumstances under which the polarization signals from different regions in the atmosphere can cancel each other out because of the differences in the magnetic field geometry.

\begin{figure*}
    \centering
    \includegraphics[width=0.9\textwidth]{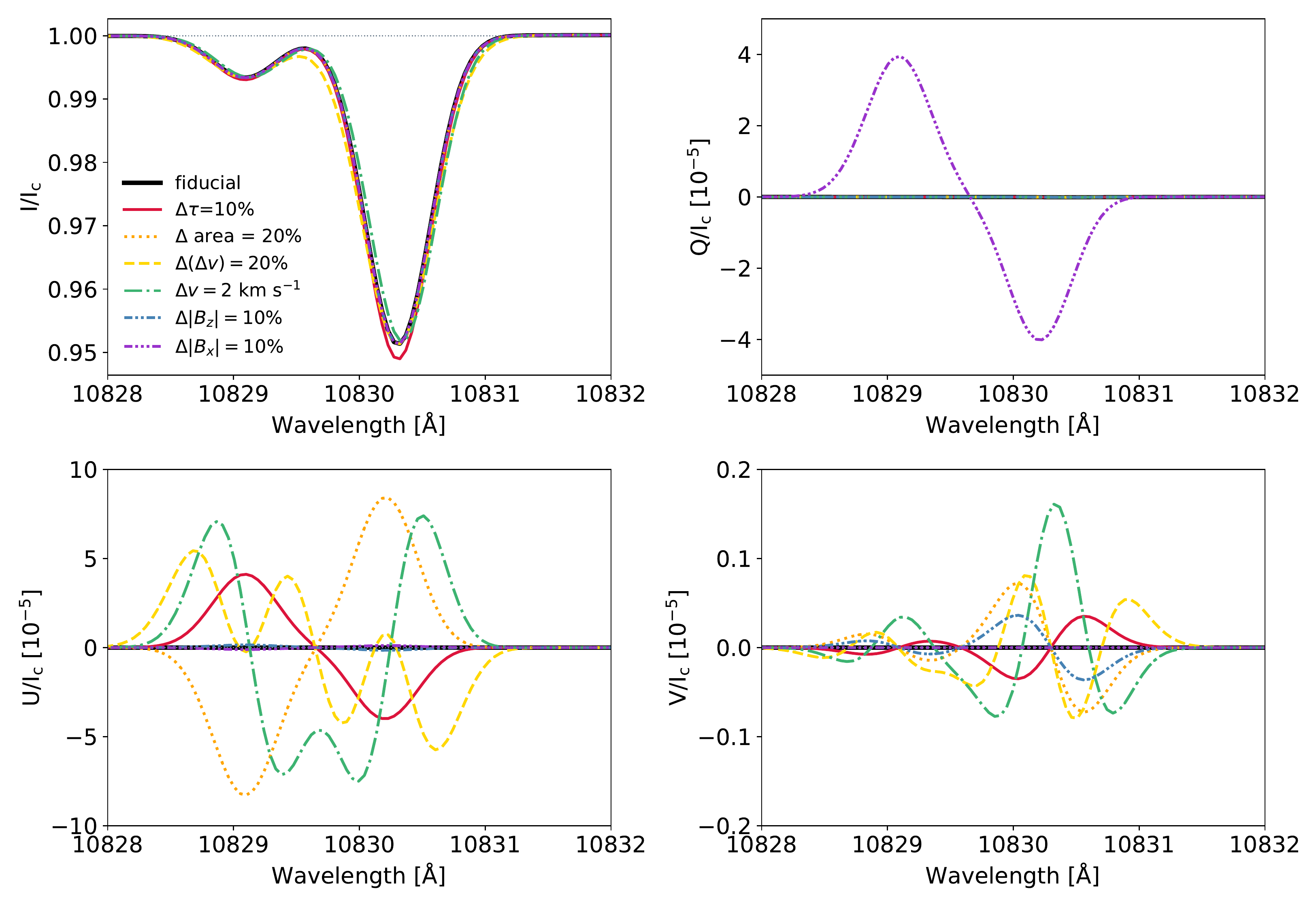}
    \caption{The fiducial case (black solid line) shows two slabs with identical properties (e.g. area, optical depth) and B-field orientation such that the polarization signals from the two slabs completely cancel out. Changing the slab properties even by a small amount, $\sim 10-20$\%, induces a linear polarization signal $\gtrsim 10^{-5}$. Cancellation of the linear polarization signal below the observability threshold of $10^{-5}$, arising from combining signals from different parts of a planetary atmosphere, hence seems unlikely.}
    \label{fig:hazel_results6}
\end{figure*}

We use \textsc{Hazel} to calculate the polarization signals arising from two independent slabs. Initially, we set all properties of the two slabs to be equal, except the magnetic field orientation. We find that the polarization signals from the two slabs almost entirely cancel out if (1) the slabs have opposite magnetic fields along the line of sight ($B_{z,1} = -B_{z,2}$) and (2) if their magnetic field components perpendicular to the line of sight are equal in magnitude, but tilted by $90^\circ$ with respect to each other (e.g. $B_{x,1} = -B_{y,2}$ and $B_{y,2} = B_{x,1}$). We show the results for one such example in \autoref{fig:hazel_results6} (`fiducial' case, black line). If the entire atmosphere could be split pair-wise into regions that mutually cancel each other, such as these two slabs, the overall polarization signal from the planetary atmosphere would be below the detection limits of current spectropolarimetric instruments. 

However, even small deviations in slab properties lead to detectable levels of linear polarization. \autoref{fig:hazel_results6} shows the polarization signal in cases when different properties of one of the slabs have been altered by a small amount ($10-20\%$) relative to the fiducial case, thereby breaking the symmetry of the problem. Even such small deviations from symmetry produce linear polarization signals on the order of a few $\times$ 10$^{-5}$, which may be reached with upcoming instruments (see \S\ref{ss:observability}).

\subsection{Observability of the Polarization Signal}
\label{ss:observability}

The polarization signals we found in this paper are small---approximately 0.1\% for a favorable field geometry and 0.01\% and weaker signals for less favorable situations---but fortunately the target stars are bright, and high signal-to-noise ratio (S/N) spectropolarimetry is possible. For spectropolarimetry dominated by source Poisson noise (the relevant regime here) and obtained by feeding the two polarization states through separate fibers to a spectrograph, the uncertainty in polarization is $\sigma(Q/I) = N_\gamma^{-1/2}$, where $N_\gamma$ is the total number of photons per bin.\footnote{Only one of the polarization Stokes parameters can be measured at a time, since this technique does not measure the correlation between the two polarization states. In an instrument such as SPIRou, the $x$ and $y$ polarizations are separated by a Wollaston prism, and rotatable quarter-wave transformers are used to map the desired Stokes parameter from the sky into Stokes $Q$ as seen by the prism \citep{2012SPIE.8446E..2EP}.} This is
\begin{eqnarray}
\sigma(Q/I) &=& 0.001 \times 10^{0.2(m_{\rm AB}-9.1)} \times \frac{3.6\,\rm m}{D}
\nonumber \\ && \times
\left( \frac{0.023}{\eta} \frac{10\,\rm km/s}{\Delta v}
\frac{10^4\,\rm s}{t_{\rm obs}} \right)^{1/2},
\label{eq:SQI}
\end{eqnarray}
where $m_{\rm AB}$ is the apparent AB magnitude of the star at $\sim 1.08\,\mu$m, $D$ is the telescope diameter, $\eta$ is the net throughput (including vignetting and fiber aperture losses), $\Delta v$ is the bin width, and $t_{\rm obs}$ is the observation time. We have scaled the instrument parameters to the SPIRou\footnote{URL: {\tt http://etc.cfht.hawaii.edu/spi/}} instrument on the Canada-France-Hawaii Telescope, and $10^4$ s is a typical single transit duration. Two of the planets with large 1083 nm transit depth are WASP-107b (depth 5.5\%; \citealt{Allart2019}) and WASP-69b (depth 3.6\%; \citealt{Nortmann2018}); their host stars have $J_{\rm AB}= 10.3$ and 8.9, respectively \citep{2003yCat.2246....0C}. HAT-P-11b, HD 209458b, and HD 189733b have lower 1083 nm transit depths, but with host stars at $J_{\rm AB} = 8.5$, 7.5, and 7.0, respectively, it should be possible to achieve smaller uncertainties on $Q/I$. All of these systems are in an appropriate declination range for CFHT/SPIRou.

\begin{figure*}
    \centering
    \includegraphics[width=0.99\textwidth]{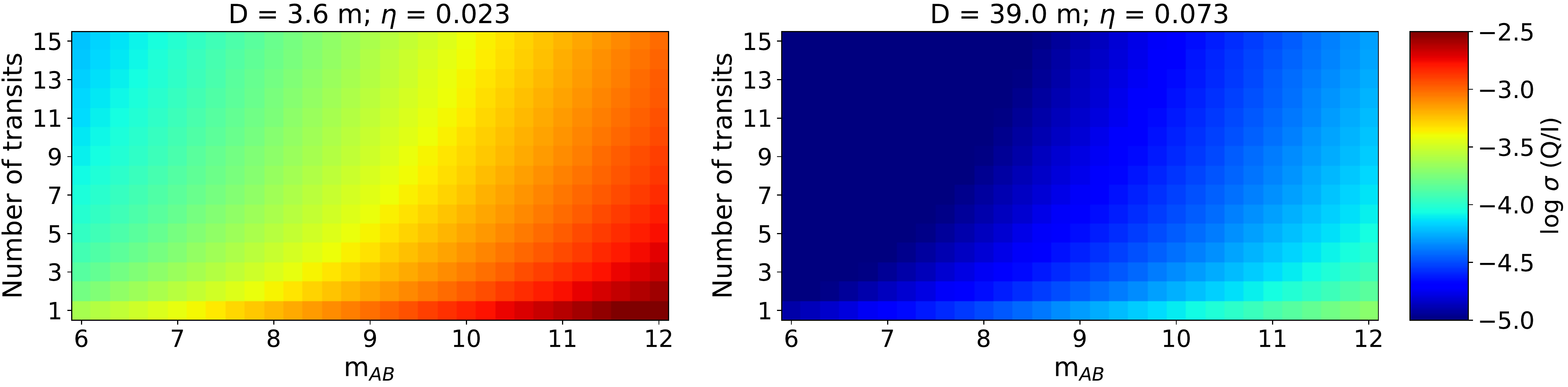}
    \caption{Uncertainty in the polarization signal (\autoref{eq:SQI}) as a function of the total exposure time (shown as the number of combined transit observations, assuming one transit lasts $10^4$~s) and the magnitude of the host star. The left (right) panel shows the results for a telescope diameter of 3.6 (39.0)~m and the instrument throughput of 0.023 (0.073), based on the corresponding values for CFHT/SPIRou (ELT/HIRES). In order to detect the polarization signal described in this work, the required uncertainty should be $\lesssim 10^{-4}$ (turquoise and blue regions). Note that this is for measurement of one polarization Stokes parameter; measurement of both $Q$ and $U$ to the same precision would require twice as many transits. 
    }
    \label{fig:observability}
\end{figure*}

With multiple transits\footnote{Here we assume that the planets are tidally locked, which should be a valid assumption for most close-in exoplanets. Therefore, the magnetic field geometry relative to the observer is expected to be similar (though, possibly not identical if it depends on stellar activity) in successive transits, which should allow combining signals from multiple transit observations.} (see \autoref{fig:observability}, left panel), and if systematics challenges associated with calibration and stellar variability can be addressed, SPIRou should enable initial exploration of the interesting parameter space for \ion{He}1 line polarization. However, if one wants to be able to detect less favorable field geometries or measure the transit phase or velocity dependence of the polarization, a future spectropolarimeter with more light-gathering capability will be required. Some improvements may be realized by optimizing materials and coatings for the $Y+J$ bands (SPIRou achieves a factor of a few higher throughput in $H+K$). The biggest improvement would be to go to an extremely large telescope. We take as an example the ELT-HIRES instrument concept \citep{2018SPIE10702E..1YM} for the Extremely Large Telescope (ELT: $D=39$ m), and based on available information set $\eta = 0.073$.\footnote{We estimated the throughput from the online exposure time calculator (ETC) for intensity mode (URL: {\tt http://www.arcetri.inaf.it/\~{}hires/etc.html}). The value of 0.073 was set to reproduce the correct S/N ratio, although it is very close to the 0.075 that one gets by multiplying the ETC input parameters since we are very source Poisson dominated.} A polarimetric mode, based on selection/splitting of the polarizations at intermediate focus, has been proposed \citep{2018SPIE10706E..1YD}. This would represent a factor of $\sim 20$ improvement in polarization sensitivity relative to SPIRou, and in principle fractional polarizations $Q/I$ down to a few$\times 10^{-5}$ would be accessible (see \autoref{fig:observability}, right panel).

\subsection{Other Sources of In-transit Polarization}

Linear polarization signals can arise during exoplanet transits from other sources, but they should not interfere with the measurements we are proposing in this study. The proposed method---high-resolution transmission spectropolarimetry---is differential in terms of both time and wavelength dependence. In other words, we propose to measure the change in polarization in-transit versus out-of-transit, as well as in the helium line versus the continuum. Therefore, any source of constant or broadband polarization should be removed in the process automatically. However, it may still contribute to the noise, which is why it is important to note that the continuum linear polarization signal of transiting exoplanets is expected to be much smaller than the signals predicted in this study.

Measuring broadband linear polarization signal during transits of exoplanets has been proposed as a method for detecting and characterizing transiting exoplanets \citep[e.g.][]{Carciofi2005, Kostogryz2015}. Searches for these signals have been conducted in recent years, without confirmed detections so far \citep{Berdyugina2008,Wiktorowicz2009, Bott2016, Bott2018}. Broadband linear polarization results from radiation scattering in the stellar atmosphere. For a spherically symmetric star, the signals integrated over the entire stellar disk cancel out; however, a transiting exoplanet breaks the symmetry of the star as seen by the observer, which results in net polarization. The expected linear polarization signal is on the order of a few $\times 10^{-6}$ at short wavelengths around 450 nm  \citep{Kostogryz2015, Kostogryz2017}. Due to the strong wavelength dependence of Rayleigh scattering, the amplitude of the polarization signal drops significantly at longer wavelengths and is expected to be negligible at 1083 nm.

Potentially more significant sources of contamination are the intrinsic stellar variability and   stellar-disk inhomogeneity in the helium triplet polarization that are due to stellar activity. The potential impact of intrinsic stellar variability in the helium line at short time-scales relevant for transit observations (a few hours) is still an open question in the context of radiation intensity, and even more so in terms of radiation polarization. Because areas of intense chromospheric absorption in the helium line are associated with active regions, and are thus unevenly distributed across the stellar disk, it is possible that a transiting planet could induce a change in the observed helium line just by occulting an exceptionally active or inactive part of the stellar disk.

Initial analyses suggest that the contamination of transmission spectra by stellar activity at 1083 nm is not very severe and should not impede observations of extended exoplanet atmospheres, at least when it comes to radiation intensity. Repeated transit observations of exoplanets at 1083 nm show consistent transit depths and similar light curves over periods of months and years, indicating that the effects of stellar activity and variability do not dominate the signal. Furthermore, by simulating exoplanet transits using synthetic spectra of F- and G-type stars with different levels of stellar activity, as well as the publicly available solar data, \citet{Cauley2018} found that the contrast between active and inactive regions at 1083 nm is small and should not result in significant contamination of helium transmission spectra. Similar investigations of the impact of stellar activity on transmission polarimetry at 1083 nm are needed, and important initial insights could be provided by studies of the spatially resolved polarization of the Sun.

\section{Summary}
\label{sec:summary}

We propose a method for directly detecting the presence of magnetic fields in the atmospheres of transiting exoplanets. The method, similar to the method originally introduced by solar physicists \citep[see][]{TrujilloBueno2002}, is based on detecting radiation polarization in the helium line triplet at 1083 nm during transits of close-in exoplanets with extended or escaping atmospheres. Using analytic and numerical calculations, we demonstrated that the presence of a transverse magnetic field induces a linear polarization signal on the order of $10^{-4}-10^{-3}$. A broad range of magnetic field strengths, including those measured for most planets in the solar system, can result in a polarization signal of this magnitude. Therefore, this method is extremely sensitive to the presence of magnetic fields in exoplanet atmospheres. Assessing the magnetic field strength, however, may be challenging in this saturated Hanle regime, unless the field strength is high enough to induce a detectable level of circular polarization, due to the longitudinal Zeeman effect. Detecting the calculated polarization signals could be achieved with future high-resolution near-infrared spectropolarimeters on large ground-based telescopes, or even with current facilities (such as SPIRou on CFHT) if multiple transit observations of the same target were combined.

\acknowledgments

We thank the anonymous referee for their comments and suggestions. We are grateful to the authors of \textsc{Hazel} for making the code publicly available. We thank Wilson Cauley, Jean-Michel D\'{e}sert, Ray Jayawardhana, Melodie Kao, Evgenya Shkolnik, and Jake Turner for useful conversations. Support for this work was provided by NASA through the NASA Hubble
Fellowship grant HST-HF2-51443.001-A awarded by the Space Telescope
Science Institute, which is operated by the Association of Universities for
Research in Astronomy, Incorporated, under NASA contract NAS5-26555. During the preparation of this work, some of the authors were supported by the Simons Foundation award 60052667 (PMC, CMH, and MS); US Department of Energy award DE-SC0019083 (PMC and CMH); and NASA award 15-WFIRST15-0008 (CMH).

\software{\textsc{Hazel} \citep{AsensioRamos2008}, matplotlib \citep{Hunter2007}, NumPy \citep{numpy}, SciPy \citep{scipy}}

%






\appendix
\section{Gas Densities Required for Depolarizing Collisions}
\label{sec:appendix}

The cross section for elastic spin-exchange (i.e., depolarizing) electron-metastable helium collisions for electron energies around 0.5 eV (typical for temperatures of $\sim 10^3-10^4$ K) is $\sigma \sim100{\,\rm\AA}^2\sim 10^{-14}$ cm$^2$ \citep{Sklarew1968}. The rate at which an electron of that energy collides with a helium atom in the metastable state, per unit volume, is therefore $R = \sigma \times \sqrt{2E/m_e} \sim 7\times 10^{-7}$ cm$^3$ s$^{-1}$. Collisions of metastable helium atoms with electrons and hydrogen atoms that cause triplet-to-singlet transitions have even lower average rates \citep[given in][section 3.3]{OklopcicHirata2018}. 
Therefore, in order for the total collision rate $n_{e} \times R$ 
to be above the scattering rate of the 1083 nm photons (at 0.05 au from a Sun-like star) of $3A_{ul} \overline{f}_{1083} = 3\times 10^7 \times 10^{-4}$ s$^{-1}$, 
the electron number density, $n_{e}$, must exceed $4\times 10^{9}$ cm$^{-3}$.

It is also possible to have elastic spin-exchange collisions between metastable helium and neutral hydrogen atoms (H{\sc\,i}) because the hydrogen atom has net electron spin. We were not able to find a published rate coefficient, but we do note that spin exchange only occurs when the He 2s and H 1s orbitals overlap each other and hence give rise to a nonzero exchange integral. Because the He 2s orbital has a wave function that declines with an $\sim 1\,$\AA\ exponential scale length, we expect the cross section at which this interaction becomes important to be of order $10 {\,\rm\AA}^2 \sim 10^{-15}$ cm$^{2}$. This would makes the He~2$^3$S\,+\,H{\sc\,i} spin exchange negligible compared to He 2$^3$S\,+\,$e^-$, since the cross section is 10 times smaller and $v$ is 40 times smaller (because of the reduced mass); thus only when the ionization fraction drops below $\sim 1/(10\times 40)\sim 0.0025$ would H{\sc\,i} collisions dominate.

According to hydrodynamic simulations of escaping exoplanet atmospheres \citep[e.g.,][]{Salz2016}, total gas densities of $\sim 10^{10}$ cm$^{-3}$ are expected at altitudes below $\sim 1.1 - 1.2$ R$_{pl}$, where most of the gas is neutral and electron collisions become further suppressed by the low ionization fraction ($\lesssim 10^{-2}$). Furthermore, observations of helium signals from exoplanets made so far indicate that the helium absorption arises in atmospheric layers extending out to much higher altitudes, up to $\sim 2-3$ R$_{pl}$ \citep{Nortmann2018, Allart2018}. The expected gas densities at these high altitudes are between $10^6$ and $10^9$ cm$^{-3}$ \citep[][with lower densities being more representative of larger planets, which are more favorable for transit observations]{Salz2016}. Because most hydrogen atoms in this region are expected to be ionized, the number density of free electrons is roughly of the same order. Hence, in most cases of interest, depolarizing electron collisions should not be able to destroy the lower-level polarization of helium atoms in these upper atmospheric layers.

\begin{figure*}
    \centering
    \includegraphics[width=0.79\textwidth]{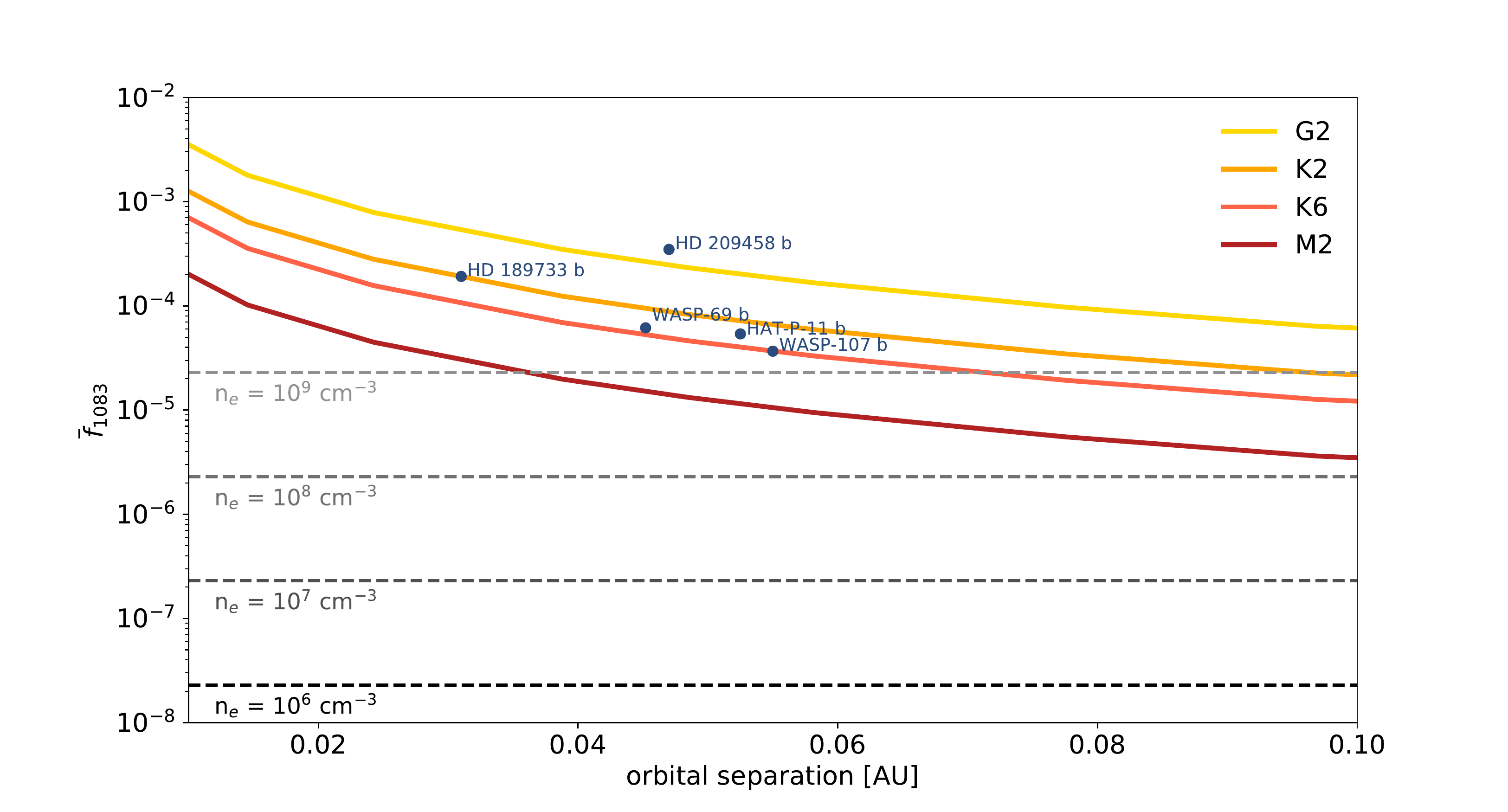}
    \caption{Solid lines show the phase space density of 1083 nm photons, $\overline{f}_{1083}$, as a function of orbital distance around stars of spectral types G2 to M2. Dashed lines show thresholds below which depolarizing collisions can destroy atomic polarization, assuming a certain electron  number density in a planetary atmosphere, $n_e$. Planets in which the helium 1083 nm absorption has been detected occupy the part of the parameter space in which depolarizing collisions do not play the dominant role, which is why we did not take them into account in our calculation.}
    \label{fig:depolarization}
\end{figure*}

However, for planets around stars of later spectral types than the Sun, the flux of 1083 nm photons at the same orbital distance can be significantly smaller. In \autoref{fig:depolarization}, we show the phase space density of 1083 nm photons as a function of planet's orbital separation for different late-type stars. The yellow curve, representing the Sun, is obtained from HAZEL, whereas the other lines are proportionally scaled down, based on the observed flux density in the continuum near the 1083 nm line of stars of given spectral type in the MUSCLES survey \citep{France2016}. Specifically, the K2 curve is modeled after $\epsilon$ Eri, K6 is based on HD 85512, and M2.5 on GJ 176. Dashed lines indicate the levels of irradiation at which the rate of scattering of 1083 nm photons becomes comparable to the rate of depolarizing collisions, at a given electron number density, according to the above-described calculation. In other words, regions of parameter space in which the solid lines lie above the dashed lines are favorable for using the Hanle effect as a probe of magnetic fields in exoplanets. This technique can only be applied up to a certain orbital distance from the host star, and this limiting distance is smaller for cooler stars and for planetary atmospheres of higher densities. As shown in \autoref{fig:depolarization}, exoplanets with reported detections of helium (indicated by blue dots) all lie well within the favorable part of the parameter space, for the expected electron densities of $n_e \sim 10^6 - 10^9$ cm$^{-3}$.

\bibliography{refs_helium_pol}
\end{document}